\documentclass[showpacs,aps]{revtex4-1} 

\usepackage{epsfig}
\usepackage{amsmath}

\begin{document}

\title{\bf General magnetized Weyl solutions: Disks and  motion of charged 
particles}


\author{Cristian H. Garc\'{\i}a-Duque } \email[e-mail: ]{garciahcristian@hotmail.com}

\author{Gonzalo Garc\'{\i}a-Reyes} \email[e-mail: ]{ggarcia@utp.edu.co}

\affiliation{Universidad  Tecnol\'ogica de Pereira, Departamento de 
F\'{\i}sica,  A. A. 97, Pereira, Colombia}




\begin{abstract}
We construct three families of  general magnetostatic axisymmetric exact solutions of Einstein-Maxwell equations  in spherical coordinates, prolate, and oblates.  The solutions obtained are then  presented in  the system of generalized spheroidal coordinates which is a generalization of the previous systems. The method used to build such   solutions is the well-known   complex potential formalism proposed by  Ernst, using as seed  solutions  vacuum solutions of  the Einstein field equations.  We show   explicitly  some particular  solutions among them a magnetized Erez-Rosen  solution and a magnetized Morgan-Morgan  solution, 
which we interpret as  the exterior gravitational field of  a finite dislike source immersed in a magnetic field. From them we also construct  using  the well known  ``displace,                                                                                                                          cut and reflect'' method  exact solutions representing relativistic  thin disks of infinite extension.
We then analyze   the  motion of electrically  charged test particles around these fields for  equatorial circular orbits and we discuss their stability  against radial perturbations. For  magnetized  Morgan-Morgan fields we find  that inside of disk  the presence of magnetic field provides the possibility of to find 
relativist charged particles  moving  in both
prograde  and retrograde direction.

\end{abstract}

\maketitle

\section{Introduction}

Magnetic fields play an important role  in the study of 
astrophysical objects such as
neutron stars, white dwarfs, pulsars, black
holes  and galaxy formation. In fact, several observations show that there are various scenarios where the magnetic fields and general 
relativity can not be neglected.
One of them is the presence of strong magnetic fields in active galactic nuclei \cite{6, 7, 8, 9}. These nuclei are known to produce more radiation than the rest of the entire galaxy and directly affect
its structure and evolution. Another scenario is the production of relativistic collimated jets in the inner regions
of accretion discs, which can be explained considering
magneto-centrifugal mechanisms \cite{10, 11, 12, 13, 14, 15}.
Also, magnetic fields are important in understanding the
interplay between magnetic and thermal processes for
strongly magnetic neutron stars \cite{4, 16, 17}. At least
10\% of all neutron stars are born as magnetars, with
magnetic fields above $10^{14}$ G \cite{18, 19, 20}. Analytical models that describe these astrophysical objects are often associated with solutions of Einstein’s equations \cite{21, 22, 23, 24, 25, 26}.
In the search for more realistic models for compact
stellar systems, the energy-momentum tensor, the source
of Einstein’s equations, is modified by introducing more
complex terms that take into account additional physical
properties as, for example, electromagnetic fields \cite{LPU}.

Stationary or static axially symmetric  exact solutions to the Einstein  field equations
representing  relativistic thin disks are of  great astrophysical importance since
they can be used  as  models for certain galaxies, accretion disks, and the superposition of a black holes and a galaxy or an accretion disk as in the case of quasars. Static thin disks without radial pressure
were first studied by Bonnor and Sackfield \cite{BS}, 
and Morgan and Morgan \cite{MM1}, and with radial  pressure by Morgan and Morgan \cite{MM2}.  Also thin disks with radial tension were considered \cite{GL1}.  
Several classes of exact solutions of the  Einstein field  equations
corresponding to static thin disks with or  without radial pressure have been
obtained by different authors
\cite{LP,LO,LEM,LL1,BLK,BLP,GE}.
Rotating  thin disks that can be considered as a source of a Kerr metric were
presented by  Bi\u{c}\'ak and  Ledvinka \cite{BL}, while rotating disks with
heat flow were were studied by Gonz\'alez and Letelier \cite{GL2}.
Disk sources for stationary axially symmetric spacetimes  with  electromagnetic fields, especially magnetic fields, 
are also of astrophysical importance in the study of neutron stars, white dwarfs and  galaxy formation. 
In such situation one has to study the coupled Einstein-Maxwell equations.
Thin disks in presence of electromagnetic field  have been discussed as sources for
Kerr-Newman fields \cite{LBZ,GG4}, magnetostatic  axisymmetric fields \cite{LET1},
conformastationary metrics \cite{KBL},  while models of electrovacuum static
counterrotating dust disks  were presented in \cite{GG1}. Charged perfect fluid
disks were also studied by Vogt and Letelier \cite{VL2}, and  charged perfect
fluid disks as sources of  static and  Taub-NUT-type spacetimes by Garc\'\i
a-Reyes and Gonz\'alez \cite{GG2,GG3}.

In all the above cases, the disks are obtained by an ``inverse problem''
approach, called by Synge the ``{\it g-method}'' \cite{SYN}. The method works
as follows: a solution of the vacuum Einstein equations is taken, such that
there is a discontinuity in the derivatives of the metric tensor on the plane
of the disk,  and the energy-momentum tensor is obtained from the Einstein
equations. The physical properties of the matter  distribution are then studied
by an analysis of the surface energy-momentum tensor so obtained. Another
approach to  generate disks is by solving the Einstein equations given a source
(energy-momentum tensor). Essentially, they are obtained by solving a
Riemann-Hilbert problem and are highly nontrivial 
\cite{NM,KLE1,KR,KLE2,FK,KLE3,KLE4}.
A review of this kind of disks solutions to the Einstein-Maxwell
equations was presented by Klein in \cite{KLE5}.

Motion of matter near compact stars and black holes has been discussed widely  in literature.  The interplay between gravitational and electromagnetic interaction is essential for characteristics of the motion, namely 
its stability properties. Motivation for these studies arises from the problem of motion and
acceleration of matter (charged particles or dust grains) \cite{Karas1, Sengupta,Felice, Karas2}.
The study of the interaction 
between particles and electromagnetic fields in curved spacetimes
is also of astrophysical interest, such is the case of  
strong synchrotron radiation emerging  galactic cores,  which can be explained admitting
the existence in those regions of extended and very intense magnetic fields, interacting with
ultrarelativistic electrons. Such magnetic fields could originate in the inner part of an accretion
disc around the central black hole \cite{Frank,Belvedere}. 
Also have been shown  that the presence of a
strong magnetic field provides the possibility of relativistic motion
of  counterrrotating matter \cite{Aliev}.

In the present work we construct three families of general magnetostatic axisymmetric exact solutions of Einstein-Maxwell equations in spherical coordinates, prolate, and oblates. The zero order of this solutions were investigated previously in the Ref. \cite{GG1},  and although the
models of thin disks constructed  there  satisfied all the energy conditions, the solutions
are not asymptotically flat.
Thus the solutions  presented  in this work  are
a generalization of  solutions discussed  in that reference, and,  except
the  zero order, the new families of solutions are asymptotically flat but with total mass  zero. Hence, all discussed solutions with magnetic field are either singular,
not asymptotically flat or massless.   
Moreover, in all cases we build models of thin disk in which all the energy conditions are satisfied,
which  characterizes a  matter distribution made of usual matter.
In the case of  oblates coordinates, the   built solutions
correspond to the  magnetized version of the Morgan-Morgan vacuum solution \cite{MM1}.
These solutions are interpreted  as  the exterior gravitational field of  a finite dislike matter distribution immersed in a magnetic field,  and  thus
can be used to model disklike astrophysical objects such as galaxies, accretion disks, and certain stars in presence of  magnetic fields. 
However, unlike the part material or the electrostatic case, we show that  the  source of the
magnetic field is not planar but of a different origin such as a remnants or fossil magnetic field \cite{FOSSIL}, or can come from
external sources, such as the presence of a nearby magnetars or 
neutron stars. On the other hand, 
even though realistic  disklike  sources have thickness, 
in first approximation these astrophysical objects  can be considered to be very thin, e.g., in our Galaxy the radius of the disk is
$10$  $kpc$ and its thickness is $1$ $kpc$.
In all cases we also construct 
relativistic models of  infinite thin disks.  We find  for
all the values of parameters that  the surface energy density decreases
rapidly  which permits  one can define a cut off radius  and, in principle,  to consider these disks as finite.  
In addition, we  analyze the  motion of charged test particles around these fields for circular equatorial orbits and  we  discuss the stability of this orbits against radial perturbations.   For  magnetized  Morgan-Morgan fields we find  that inside of disk  the presence of magnetic field provides the possibility of to find 
relativist charged particles  moving  in both
prograde  and retrograde direction.  Even though this
result can be seen as merely theoretical, there are observational
evidence of  counterrotating matter components  in certain types of galaxies
\cite{RGK,RFF,BER,STRUCK,CBG}.

The paper is organized as follows.  In section II  we make a review of the Ernst's method in the case of a magnetostatic axisymmetric 
spacetime and we construct three  families of  solutions of Einstein-Maxwell equations in spherical coordinate, prolate and oblate, and then in generalized spheroidal coordinates which are a generalization of the previous cases.  We also show   explicitly  some particular  solutions  among them a  magnetized Erez-Rosen  solution and magnetized Morgan-Morgan  solution,  and we analyze limiting cases.

In section III  we present a
summary of the procedure to obtain  models of  thin disks  with a
purely azimuthal pressure  and  currents for the  Einstein-Maxwell equations. In particular, we obtain expressions for the surface 
energy-momentum tensor  and the surface current density of the disks.   
In the case of the magnetized  Morgan-Morgan  solutions, these are interpreted  as  the exterior gravitational field of  a finite dislike source immersed in a magnetic field.  Then we construct  using  the well known ``displace,
cut and reflect'' method  exact solutions representing relativistic  thin disks of infinite extension.

In  section IV  we   study for some  particular solutions   the   equatorial circular motion of charged test particles and  we also discuss their stability against radial perturbation  
using  an extension of Rayleigh  criteria of stability \cite{RAYL,FLU,LETSTAB}. In particular we analyze the circular
velocity and the specific angular momentum of the particles. 
In order to compare the behavior of these 
physical quantities  with other  known  magnetized solutions,  we 
also study the electrogeodesic motion of test particles  and  their stability   for a  Kerr-type solution  or magnetic dipole solution
\cite{BONNOR}.
Finally, in Section V we summarize and discuss the results obtained.


\section{General magnetized Weyl solutions}

The  simplest metric  to
describe a  static axially symmetric spacetime is the Weyl's line element \cite{KRAMER}
\begin{equation}
ds^2 = - \ e^{2 \psi} dt^2 \ + \ e^{- 2 \psi} [\rho^2 d\varphi^2 + e^{2 \Lambda}
(d\rho^2 + dz^2)] , \label{eq:met}
\end{equation}
where  $(t, \varphi,\rho, z)$ are the Weyl canonical coordinates, and  $\psi$ and $\Lambda$  are functions of the coordinates $\rho$ and $z$ only.  For the
coordinates we also use the notation 
$(x^0, x^1,x^2,x^3)=(t, \varphi,\rho,z)$.  The vacuum
Einstein-Maxwell equations, in  geometrized units such that 
$ G = c  = 1$,  are given by 
\begin{subequations}
\begin{eqnarray}
&   &    R_{ab} \  =  \ 8 \pi T_{ab}, \label{eq:einmax1} \\
&   &     \nonumber       \\
&   &     T_{ab} \  =  \ \frac{1}{4 \pi} \left [ F_{ac}F_b^{ \ c} - \frac 14 g_{ab}F_{cd}F^{cd} \right ], \label{eq:tab} \\
&   &     \nonumber    \\
&   &    F^{ab}_{ \ \ \ ; b} = 0, \label{eq:einmax2}   \\
&   &     \nonumber       \\
&   &   F_{ab} =  A_{b,a} -  A_{a,b},
\end{eqnarray}\label{eq:einmax}\end{subequations} 
where  $T_{ab}$ is the electromagnetic energy-momentum  tensor, $F_{ab}$ is the
electromagnetic field tensor, and $A_a=(\phi,A,0,0)$ is the four-potential, where
$\phi$ is the electric potential and $A$ the magnetic potential which are also
functions of $r$ and $z$ only.  Further,  $( \ )_{,a}=\partial /\partial
x^a$,  and $( \ )_{;a}$ means covariant derivate.

For the metric (\ref{eq:met}) and in magnetostatic case,  the Einstein-Maxwell equations are
\begin{subequations}\begin{eqnarray}
\nabla\cdot[\rho^{-2}f\nabla A]&=&0, \label{eq:e-m1} \\
f\nabla^{2}f&=&\nabla f\cdot\nabla f + 2\rho^{-2}f^{3}\nabla A\cdot\nabla A, \label{eq:e-m2} \\
\Lambda,_\rho&=&\rho\left(\psi^{2},_{\rho}-\psi^{2},_{z}\right)+\frac{1}{\rho}\left(A^{2},_{\rho}-A^{2},_z\right)f, \label{eq:lambda1} \\
\Lambda,_z&=&2\rho\psi,_{\rho} \psi,_{z}+\frac{2}{\rho}A,_{\rho}A,_{z}f, \label{eq:fylambda2}
\end{eqnarray}\end{subequations}
where $f=e^{2\psi}$. The equations (\ref{eq:e-m1}) y (\ref{eq:e-m2}) are equivalent to
\cite{E1,E2}
\begin{subequations}
\begin{eqnarray}
f \Delta {\cal E} &=& (\nabla {\cal E} + 2\Phi^\ast\nabla\Phi) \cdot
\nabla{\cal E}, \label{eq:ece1}     \\
&  & \nonumber  \\
f \Delta \Phi &=& (\nabla {\cal E} + 2\Phi^\ast\nabla\Phi) \cdot \nabla\Phi,
\label{eq:ece2} 
\end{eqnarray}\label{eq:ece}\end{subequations}
where  $\Delta$ and  $\nabla$ are  the  standard differential operators in cylindrical coordinates, $f=
e^{2\psi}$, and ${\cal E}$ and $\Phi$ are complex potentials which 
in the case static   ${\cal E}={\cal E}^*$ and  for the case 
magnetostatic  $\Phi ^* = -\Phi$ .  The above equations are called  Ernst equations. The metric functions are obtained via 
\begin{subequations}\begin{eqnarray}
f  &=& {\cal E} +\Phi \Phi^*  ,  \\
&&	\nonumber	\\
\Lambda _{,\zeta} &=&  \frac{\sqrt{2} \rho}{4 f^2} ( {\cal E}_{,\zeta} +2 \Phi ^*
\Phi_{,\zeta} )({\cal E}_{,\zeta} +2 \Phi  \Phi^*_{,\zeta} ) 
 - \ \frac{\sqrt{2} \rho}{f}\Phi_{,\zeta}\Phi^*_{,\zeta} ,
\label{eq:Lam} 
\end{eqnarray}\end{subequations}
where $\sqrt 2\zeta = \rho +iz$, so that $\sqrt 2 \partial_{,\zeta}=
\partial_{,\rho}-i\partial_{,z}$, and  the magnetic potential $A$ is  related to  $\Phi$ via
\begin{equation}
A_{,\zeta} =   i \frac \rho f ( {\rm Im} \Phi )_{,\zeta}.
\end{equation}
 
Taking ${\cal E}$ as function of $\Phi$,  from (\ref{eq:ece1}) and 
(\ref{eq:ece2}) it follows that
\begin{equation}
( {\text Re} {\cal E} + |\Phi|^2 ) \frac{d^2 {\cal E} }{d \Phi ^2} \nabla \Phi \cdot
\nabla \Phi = 0,
\end{equation}
and hence that ${\cal E}$ is a lineal function of $\Phi$. Using the
boundary conditions ${\cal E} \rightarrow 1$ and $\Phi \rightarrow 0$ at infinity, we obtain
\begin{equation}
{\cal E} = 1 - 2q^{ - 1} \Phi , \label{eq:calE}
\end{equation}
where  $q$ is a complex constant. With the  change of variable  
\begin {equation}	
{\cal E} = \frac{\xi - 1}{\xi + 1}, 
\end{equation}
then  (\ref{eq:calE}) implies
\begin{equation}
\Phi  = \frac{q}{ \left( \xi  + 1 \right) } ,
\end{equation}
and the Ernst equations read  
\begin{equation}
[\xi\xi^{\ast} - (1 - qq^{\ast})]\nabla^{2}\xi = 2\xi^{\ast}\nabla\xi\cdot\nabla\xi . \label{eq:ernstvar}
\end{equation}
Then making  $\xi=(1-qq^{\ast})^{1/2}\hat  \xi$ the  equation 
(\ref{eq:ernstvar}) takes of form 
\begin{equation}
	(\hat  \xi\hat  \xi^{\ast} - 1)\nabla^{2}\hat  \xi = 2\hat  \xi^{\ast}\nabla\hat  \xi\cdot\nabla\hat  \xi,
\end{equation}
which is the Ernst equation in the vacuum \cite{E2}. So given a solution of the Einstein field equations in vacuum 
$ \hat \xi $ (seed solution) we can construct a solution  of the
Einstein-Maxwell field equations.
A solutions for this equations is 
 \begin{equation}
	\hat \xi = - e^{i\alpha}\coth \hat \psi,
\end{equation}
where the function $\hat  \psi$ satisfies the Laplace's equation 
\begin{equation}
	\nabla^{2} \hat \psi = 0.  \label{eq:laplace}
\end{equation}

The case $\alpha = 0$ corresponds
to the well-known Weyl vacuum solutions and the magnetized solutions  (taking  $\Phi$ imaginary)    built from
them are the magnetized Weyl solutions. The metric functions and magnetic potential are given by  
\begin{subequations}\begin{eqnarray}
 f &=& \frac{4}{{\left[ {\left( {1+a } \right)e^{ - \hat  \psi }  
+ \left( {1-a} \right)e^{\hat  \psi } } \right]^2 }}, \label{eq:f} \\
\Lambda &=& \hat \Lambda,  \label{eq:lambda0}   \\ 
 A_{,\rho }  &=&  - b\rho \hat \psi _{,z},  \label{eq:Arho}  \\
A_{,z}  &=& b\rho \hat \psi _{,\rho } , \label{eq:Az} 
\end{eqnarray}\end{subequations}
where  $a=\sqrt{1+b^2}$, being  $b$ is the parameter that controls the
magnetic field,  and  $\hat \Lambda$ is the metric potential $\Lambda$ corresponding to the seed solution, that is taking  ${\cal E} = e^{2\hat  \psi}$ and $\Phi = 0$. 

In terms of generalized  spheroidal coordinates
 ($\xi$,$\eta$) \cite{RAMOS} which are related to Weyl
coordinates ($\rho$,$z$) by
\begin{subequations}\begin{eqnarray}
\rho ^2 & =& k^2 (\xi ^2  - \sigma ^2 )\left( {1 - \eta ^2 } \right), \\
z &=& k\xi \eta,
\end{eqnarray}\end{subequations}
with $k$ and $\sigma$ constants, the equations for the magnetic potential  (\ref{eq:Arho}) -  (\ref{eq:Az}) can be cast as
\begin{subequations}\begin{eqnarray}
 A,_\xi   &=&  - kb\left( {1 - \eta ^2 } \right)\hat  \psi ,_\eta \label{EQ:Aeta} ,  \label{eq:Agen1} \\ 
 A,_\eta   &=& kb\left( {\xi ^2  - \sigma^2 } \right)\hat  \psi ,_\xi  \label{eq:Agen2}.
\end{eqnarray}\end{subequations}
and for the function $\Lambda$ 
\begin{subequations}\begin{eqnarray}
 \Lambda  ,_\xi   = \left( {\frac{{1 - \eta ^2 }}{{\xi ^2  - \sigma ^2 \eta ^2 }}} \right)\left[ {\xi \left( {\xi ^2  - \sigma ^2 } \right)\hat  \psi ,_\xi  ^2  - \xi \left( {1 - \eta ^2 } \right)\hat  \psi ,_\eta ^2  - 2\eta \left( {\xi ^2  - \sigma ^2 } \right)\hat  \psi ,_\xi  \hat  \psi ,_\eta  } \right], \label{eq:Lambgen1}  \nonumber \\ 
  \\
\Lambda  ,_\eta   = \left( {\frac{{\xi ^2  - \sigma ^2 }}{{\xi ^2  - \sigma ^2 \eta ^2 }}} \right)\left[ {\eta \left( {\xi ^2  - \sigma ^2 } \right)\hat  \psi ,_\xi  ^2  - \eta \left( {1 - \eta ^2 } \right)\hat  \psi ,_\eta  ^2  + 2\xi \left( {1 - \eta ^2 } \right)\hat  \psi ,_\xi  \hat  \psi ,_\eta  } \right] .\nonumber   \label{eq:Lambgen2}  \\
 \end{eqnarray}\end{subequations}
When  $\sigma = 1$ we have the prolate coordinates,  when 
$\sigma = i$ we have the oblate coordinates, and the case
$\sigma =0$ with   $\eta = \cos \theta$ and  $k=1$ corresponds to
the spherical coordinates. As $\hat  \psi$ satisfies the Laplace's equation, the integrability of the systems (\ref{eq:Agen1}) - (\ref{eq:Lambgen2})  is guaranteed. So  we can choose any of these  equations for find $A$ or  $\Lambda$.


\subsection{Solutions in spherical coordinates}

In terms of  spherical coordinates ($r$,$\theta$) which are related to  Weyl coordinates ($\rho$, $z$) by
\begin{equation}
\rho = r \sin \theta, \ \ \ \ z = r \cos \theta, \label{eq:cooresfe}
\end{equation}
with $0 \leq r \leq \infty $ and $- \pi \leq \theta \leq \pi$,
the asymptotically flat  general solution of Laplace's equation (\ref{eq:laplace}) can be written as
\begin{equation}
\hat  \psi  = - \sum\limits_{n = 0}^\infty  {c_n \frac{{P_n \left( {\cos \theta } \right)}}{{r^{n + 1} }}} , \label{EQ:seriesfe}
\end{equation}
where  $c_n$ are constants and  $P_n ( {\cos \theta })$ are the Legendre polynomials. The magnetic
potential obtains from of any  the equations (\ref{eq:Agen1}) - (\ref{eq:Agen2})  in spherical coordinates
\begin{equation}
A= \int_{\pi}^{\theta} A_{, \theta} d  \theta  =  - b \int_{\pi} ^{\theta} r^2  \sin  \theta \hat  \psi ,_r d  \theta ,
\end{equation}
where the integral limits are chosen by requiring that the function $ A $
to be regular on the axis of symmetry. So
\begin{eqnarray}
A &=& - b\sum\limits_{n = 0}^\infty  {c_n (n + 1)\frac{1}{{r^n }}} \int_{\pi} ^{\theta}  \sin \theta P_n ( \cos  \theta ) 
d \theta \nonumber \\
&=&  b\sum\limits_{n = 0}^\infty  {c_n (n + 1)\frac{1}{{r^n }}} \int_{\pi }^{\theta}   P_n ( \cos  \theta ) 
d (\cos  \theta). \nonumber 
\end{eqnarray}

With the change of variable $y=\cos \theta $ the above expression takes the form
\begin{equation}
A=  b\sum\limits_{n = 0}^\infty  {c_n (n + 1)\frac{1}{{r^n }}} \int_{-1} ^ {y}  { P_n \left( {y} \right)} 
d y \nonumber 
\end{equation}
and using the identity \cite{BATEMAN}
\begin{equation}
\int_{-1}^{y} P_n(y)dy = \frac{1}{2n+1} \left( P_{n+1}(y) - 
 P_{n-1}(y) \right),
\ \ n \geq 0, \ \ P_{-1}= -1, \label{EQ:iden1}
\end{equation}
we obtain
\begin{equation}
A =   b\sum\limits_{n = 0}^\infty  {\frac{{c_n (n+1) }}{{ 
(2n+1)r^n }}} \left[ { P_{n+1} (\cos \theta ) - P_{n - 1} (\cos \theta )} \right].
\end{equation}

Finally, using the identity \cite{ARFKEN}
\begin{equation}
P_{n+1} (\cos \theta ) - P_{n - 1} (\cos \theta )
= \frac{2n+1}{n+1} \left (\cos \theta P_n(\cos \theta) - 
 P_{n-1}(\cos \theta)  \right ), \label{EQ:iden2}
\end{equation}
one finds  that  the magnetic potential can be written as 
\begin{equation}
A =   b\sum\limits_{n = 0}^\infty  {\frac{{c_n }}{{r^n }}} \left[ {\cos \theta P_n (\cos \theta ) - P_{n - 1} (\cos \theta )} \right]. 
\label{eq:esfeA}
\end{equation}

The same above procedure is carried out to find 
$ \Lambda $ \cite{KRAMER}  and we obtain 
\begin{eqnarray}
\Lambda &=&\int_{\pi}^{\theta} \hat \Lambda ,_\theta {d\theta} \nonumber \\
&=&    - \sum\limits_{l,m = 0}^\infty  {\frac{{c_l c_m \left( {l + 1} \right)\left( {m + 1} \right)}}{{\left( {l + m + 2} \right)r^{l + m + 2} }}\left[ {P_l P_m  - P_{l + 1} P_{m + 1} } \right]}.
\end{eqnarray}

Since $r=\sqrt{\rho^2 + z^2}$ the solutions have only a true singularity
at the point $\rho=0$, $z=0$. Hence the solutions are regular on the axis of symmetry, except at the origin. Such as  singularity could  be 
interpreted as the inside some kind of  black hole and the  solutions as  the   gravitational field produced by such source. 
The Curzon-Scott-Szekeres  black holes \cite{Scott1,Scott2}  provides an example of this type of spacetimes. However, the problems associated with the complicated structure of the singularity
in these space-times can be avoided by considering the  solution
to represent the exterior field of some finite source, e.g, representing
the field of a disk (section III).
In addition, since $|\cos(\theta)| \leq 1  $ and  $|P_n(\cos \theta)| \leq 1$ 
\cite{ARFKEN} the magnetic potential A (\ref{eq:esfeA}) tends to
zero at infinity only for $ n \geq 1 $. Hence, the solutions are   asymptotically flat  only for $ n \geq 1 $, i.e., if the coefficient
$c_0$ vanishes.  On the other hand, 
for the line element (\ref{eq:met}), in terms of the spherical coordinates 
(\ref{eq:cooresfe}),
\begin{equation}
ds^2 = - \ e^{2 \psi} dt^2 \ + \ e^{- 2 \psi} [r^2 \sin^2 \theta d\varphi^2 + e^{2 \Lambda} (dr^2 + r^2d \theta^2)] 
\end{equation}
the total mass 	\cite{Bardeen}
\begin{equation}
m = - \frac{1}{4 \pi} \lim _{r \rightarrow \infty} \int _{S_r} K^{a;b} d S_{ab}
\label{eq:masa}
\end{equation}
(where $S_r$ is a coordinate sphere with radius $r$ and $K = \partial t$ is the static Killing vector) for an asymptotically flat solution
is given by
\begin{equation}
m = -  \lim _{r \rightarrow \infty} (r^2 \psi_{,r}).
\end{equation}
For a seed solution of the form (\ref{EQ:seriesfe}) we have 
\begin{equation}
\hat m = -  \lim _{r \rightarrow \infty} (r^2 \hat \psi_{,r}) = -c_0. 
\label{eq:msemilla}
\end{equation}
From (\ref{eq:f})
\begin{equation}
\psi_{,r}=  \hat \psi_{,r}  \left[ \frac { (1+a){\rm e}^{- \hat \psi} - (1-a) {\rm e}^{\hat \psi } }
{ (1+a){\rm e}^{-\hat \psi} + (1-a) {\rm e}^{\hat \psi } }  \right ], 
\end{equation}
so that  using (\ref{eq:msemilla}) and the fact that $\psi \rightarrow 0$  for $r \rightarrow  \infty$ we obtain
\begin{equation}
m = -  \lim _{r \rightarrow \infty} (r^2 \hat \psi_{,r})  \lim _{r \rightarrow \infty}   \left[ \frac { (1+a){\rm e}^{- \hat \psi} - (1-a) {\rm e}^{\hat \psi } }
{ (1+a){\rm e}^{-\hat \psi} + (1-a) {\rm e}^{\hat \psi } }  \right ]
= \hat m a = -a c_0.
\end{equation}
 
Now if one chooses $c_0 = 0$ in order to guarantee asymptotical flatness this leads to $m = 0$ for arbitrary
coefficients $c_1$, $c_2$, etc. These parameters contribute
to higher multipole moments \cite{QUEVEDO}  but not to the mass. 
According to the positive mass theorem this imply that the spacetime must be singular or Minkowski space.
So when the magnetic field is present, i.e., if $b \neq 0$, these solutions are massless in the asymptotically flat case.
Moreover, since the expression for the mass 
(\ref{eq:masa})  is invariant, this result is true for any
coordinate system.  Therefore, in presence of magnetic field these solutions  are either singular, not asymptotically flat or massless.  

For  $n=0$ we have that 
\begin{equation}
\hat  \psi = - \frac{m}{r},
\end{equation}
where we have chosen $ c_0 = m $. Therefore, the exact solution of Einstein-Maxwell field equations  for $ n = 0 $ is
\begin{subequations}\begin{eqnarray}
e^{\psi} & = & \frac{2}{ (1+ a )e^{ m/r } + ( 1-a)e^{- m/r } }, \\
\Lambda &=& -\frac{m^2}{2 r^2}  \sin^2 \theta,   \\
A &=& bm(\cos \theta + 1 ).
\end{eqnarray}\end{subequations}

This solution is the magnetized version of the Chazy-Curson vacuum solution \cite{CH,C,GG1}. The solution is regular on the axis of symmetry but not is asymptotically flat. Therefore, this solution represents the exterior gravitational field of  a source of finite mass in the presence of a magnetic field spread throughout the space (infinite).

We now consider  the sum of the first and second terms $n=0$ and  $n=2$
of series (\ref{EQ:seriesfe}). In this case we have 
\begin{equation}
\hat  \psi  = -\frac{{c_0 }}{{r }} - \frac {c_2} {2r^3} (3 \cos ^2 \theta  - 1),
\end{equation}
and the general solutions in spherical coordinates in this case is
\begin{subequations}\begin{eqnarray}
A &=& b c_0 (\cos \theta + 1) - \frac{3}{2} bc_2\frac{1}{{r^2 }}\cos \theta \sin ^2 \theta,  \\
\Lambda &=&  
-\frac{1}{8 r^6}\sin^2 \theta  \left [ 3 c_2 \cos ^2 \theta
\left (25 c_2 \cos^2 \theta +10 c_0r^2 -14 c_2 \right ) \right.  \nonumber \\
&& \left . + 4c_0^2r^4-6c_0c_2r^2 +3c_2^2 \right ], \nonumber \\
&&
\end{eqnarray}\end{subequations}
and  $\psi$ is given by  (\ref{eq:met}).


\subsection{ Solutions in prolate spheroidal coordinates}

In terms of the prolate coordinates ($x$, $y$) which are related to 
 Weyl coordinates
($\rho$, $z$) by
\begin{subequations}\begin{eqnarray}
\rho ^2 &=& k^2 (x^2  - 1)(1 - y^2 ), \\
z &=& kxy,
\end{eqnarray}\end{subequations}
with $x \geq 1$,  $-1 <y<1$, and $k$  a constant,
the asymptotically flat general solution of Laplace's equation (\ref{eq:laplace}) can be written as
\begin{equation}
\hat  \psi  = - \sum\limits_{n = 0}^\infty  {c_n Q_n (x)P_n (y)}, 
\label{EQ:serieprola}
\end{equation}
where  $c_n$ are constants and the  $Q_n (x)$  are the Legendre functions of the second kind
\begin{equation}
Q_n(x)= \sum _ {s=0} ^ {\infty} \frac {2^n  (n+s)! (n+2s)!}{ s!(2n + 2s +1)!} x^{-2s-n-1}. \label{EQ:qnx}
\end{equation}

From  (\ref{eq:Agen1}) - (\ref{eq:Agen2}) in prolate coordinates, we obtain 
\begin{equation}
A = -kb\left( {x^2  - 1} \right)\sum\limits_{n = 0}^\infty  {c_n Q'_n (x)\int\limits_{ - 1}^y {P_n (y)dy} } ,
\end{equation}
where the integral limits are chosen by requiring that the function $ A $
to be regular on the axis of symmetry. Using the identities  (\ref {EQ:iden1}) and  (\ref{EQ:iden2}), we have that 
\begin{equation}
A = - kb\left( {x^2  - 1} \right)\sum\limits_{n = 0}^\infty  {\frac{{c_n }}{{n + 1}}Q'_n (x)\left[ {yP_{n} (y) - P_{n - 1} (y)} \right]}.
\label{eq:prolaA}
\end{equation}

Since $Q_n(x)$ diverges for $x \rightarrow 1$, the
resulting solutions  are singular at $x=1$, corresponding to $\rho = 0$,
$-1 <y<1$, i.e., on a part of the symmetric. In the vacuum these solutions  have been interpreted  by 
several authors  as distorted static black holes \cite{Chandra}.  Using the definition for  $Q_n(x)$ (\ref{EQ:qnx}), the recurrence relation
\begin{equation}
(x^2 -1) Q'_n(x) = (n+1)[ Q_{n+1}(x) -x Q_n (x) ]
\end{equation}
and the fact that $Q_n(\infty)=0$ we find that in the infinity  
($x \rightarrow \infty$) 
\begin{equation}
\lim_{x \rightarrow \infty } (x^2 -1) Q'_n(x) =  \left\{ \begin{array}{ll}
			-1, 		& n = 0 	\\
			 0, 	&     n \geq 1  
				\end{array} \right.	.
\end{equation}
Therefore, the magnetic potential A (\ref{eq:prolaA}) tends to
zero at infinity,
\begin{equation} 
\lim_{x \rightarrow \infty } A = 0 ,
\end{equation} 
only for $ n \geq 1 $. Hence, the asymptotically flat solutions are      those for which  $ n \geq 1 $, i.e., if the coefficient
$c_0$ vanishes.  Therefore, as was discussed above, in presence of magnetic field
these solutions are either singular, not asymptotically flat or massless.

For $n=0$  we have 
\begin{equation}
\hat  \psi  = \frac {c_0}{2} \ln \left( {\frac{{x-1}}{{x + 1}}} \right). 
\end{equation}
and the magnetic potential is 
\begin{eqnarray}
A &=& -kbc_0 \left( {x^2  - 1} \right)Q'_0 (x)\left[ {y P_0 (y) - P_{ - 1} (y)} \right],\nonumber \\
    &=& -kbc_0 \left( {x^2  - 1} \right)\left( {y + 1} \right)Q'_0 (x).
\end{eqnarray}
 Using 
\begin{eqnarray}\label{EQ:Q}
 Q_0 (x) &=& \frac{1}{2}{\mathop{\rm \ln}\nolimits} \left( {\frac{{x + 1}}{{x - 1}}} \right) ,\nonumber \\ 
 Q'_0 (x) &=&  - \frac{1}{{\left( {x^2  - 1} \right)}}, 
\label{EQ:Qprima} 
\end{eqnarray}
we  obtain  
\begin{equation}
A =   kbc_0 (y + 1).
\end{equation}

So the solution in prolate spheroidal coordinates for 
$n=0$ is
\begin{subequations}\begin{eqnarray}
e^{\psi} & = & \frac{ 2(x^2-1)^{c_0/2} }{ (1+a)(x+1)^{c_0} 
+ (1-a)(x-1)^{c_0 } }, \\
\Lambda & = &  \frac{c_0^2}{2} \ln \left[ \frac{x^2-1}{x^2-y^2} 
\right ], \\
A & = & kbc_0 (y+1).
\end{eqnarray}\end{subequations}

This solution is the  magnetized version of the  Zipoy-Voorhees vacuum solution \cite{Z,V,GG1}. For $ b = 0 $ and $ c_0 = 1 $ we have the Schwarzschild solution \cite{Sch} and the case $ b = 0 $ and $ c_0 = 2 $ corresponds to the metric of Darmois \cite{KRAMER}. The solution is regular on the axis of symmetry but is not asymptotically flat. Therefore, this solution also represents the gravitational field outside
a source of finite mass in the presence of a magnetic field
spread throughout the space (infinite).

For the terms  $n=0$ and  $n=2$ we have that 
\begin{equation}
\hat  \psi = c_0\frac{1}{2} \ln \left( \frac{x-1}{x+1} \right) + \frac{1}{2}c_2 \left( 3y^2  - 1 \right)\left[ \frac 1 4  (3x^2  - 1) \ln \left( \frac{x-1}{x+1} \right) + \frac{3}{2}x  \right],
\end{equation}
and for the second term of the summation   $n=2$ the magnetic potential is
\begin{eqnarray}
 A(n=2) &=&  - kb\left( {x^2  - 1} \right)c_2 \frac{1}{2}y\left( {1 - y^2 } \right)Q'_2 (x) ,\\ 
     &=&  - \frac{1}{4}kbc_2 y\left( {1 - y^2 } \right)\left[ {3x\left( {x^2  - 1} \right) \ln \left( {\frac{{x - 1}}{{x + 1}}} \right) + 6x^2  - 4} \right]. \nonumber \\
&&
\end{eqnarray}
Therefore the solution in prolate spheroidal coordinate in this case is 
\begin{subequations}\begin{eqnarray}
A &=&  kb c_ 0 \left( {y + 1} \right)  - \frac{1}{4} kbc_2 y\left( {1 - y^2 } \right)\left[ {3x\left( {x^2  - 1} \right) \ln \left( {\frac{{x - 1}}{{x + 1}}} \right) + 6x^2  - 4} \right] , \nonumber \\ 
&&  \label{eq:erez-rosen1} \\
\Lambda &=& 
\frac {9}{64} c_ 2^2(x^2-1)(y^2-1)(9 x^2 y^2-x^2+1-y^2) 
\left [\ln \left ( \frac{x-1}{x+1} \right ) \right]^2  \nonumber \\
&& + \frac {3}{16} c_2 x(y^2-1)(27c_2 x^2 y^2-3c_2x^2-21c_2y^2+5c_2+8c_0)
\ln \left ( \frac{x-1}{x+1} \right )  \nonumber \\
& & + \frac {1}{2}(c_0+c_2)^2 \ln \left ( \frac{x^2-1}{x^2-y^2} \right )  \nonumber \\
& & +\frac {3}{16}c_2(y^2-1)(-12c_2y^2+27c_2 x^2y^2 -3c_2x^2+4c_2+16c_0),
\label{eq:erez-rosen2}
\end{eqnarray}\end{subequations} 
and $\psi$ is given by (\ref{eq:f}).
In the absence of magnetic field $ b = $ 0 and $c_0=1$, we have  the Erez-rosen 
vacuum  metric  \cite{EREZ}  so that  the case
$b \neq0$  corresponds to a magnetized Erez-rosen metric.


\subsection{Solutions in oblate spheroidal coordinates}

In terms of the oblate coordinates ($u$, $v$) which are related to  Weyl coordinates ($\rho$, $z$) by
\begin{subequations}\begin{eqnarray}
 \rho ^2 & =& k ^2 (u^2  + 1)(1 - v^2 ) ,\\ 
 z &=& k uv,
\end{eqnarray}\end{subequations}
with $u \geq 0$, $-1< v <1$, and  $k$  a constant, 
the asymptotically flat general solution of Laplace's equation (\ref{eq:laplace}) can be written as 
\begin{equation}
\hat  \psi  = - \sum\limits_{n = 0}^\infty  {c_n q_n (u)P_n (v)} , 
\label{EQ:serieobla}
\end{equation}
where  $c_n$ are constants and
\begin{equation}
q_n (u) = i^{n + 1} Q_n (iu).
\end{equation}
Again requiring that the solution to be regular on the axis of
symmetry, it follows that  
\begin{eqnarray}
A &=& - kb\left( {u^2  + 1} \right)\sum\limits_{n = 0}^\infty  {c_n q'_n (u)\int\limits_{ - 1}^v {P_n (v)dv} } ,\nonumber \\
    &=& - kb\left( {u^2  + 1} \right)\sum\limits_{n = 0}^\infty  {\frac{{c_n }}{{2n + 1}}q'_n (u)\left[ {P_{n + 1} (v) - P_{n - 1} (v)} \right]},
\end{eqnarray}
and using the identities  (\ref{EQ:iden1})  y (\ref{EQ:iden2}) we obtain
the following expression for the magnetic potential 
\begin{equation}
 A =- kb\left( {u^2  + 1} \right)\sum\limits_{n = 0}^\infty  {\frac{{c_n }}{{n + 1}}q'_n (u)\left[ {v P_{n } (v) - P_{n - 1} (v)} \right]},
\end{equation}
and again $\psi$ is given by  (\ref{eq:f}). In the absence of magnetic field $b=0$ and for $n$ even this solutions correspond to  the Morgan-Morgan vacuum solutions \cite{MM1} which   represent  the
exterior gravitational field produced by  a finite disklike source.
So  we can call   above solutions   magnetized Morgan-Morgan  solutions.  Similarly to the prolate case, the asymptotically flat solutions are  those for which  $ n \geq 1 $, i.e., if the coefficient
$c_0$ vanishes. Hence, as was discussed above, in presence of magnetic field
these solutions are either singular, not asymptotically flat or massless.

For $n=0$ we have 
\begin{equation}
\hat  \psi  = -  c_0 \cot^{ - 1} (u),
\end{equation}
and the potential magnetic is given by 
\begin{equation}
A = - kbc_0 \left( {u^2  + 1} \right)q'_0 (u)\left( {v + 1} \right),
\end{equation}
with 
\begin{eqnarray}
 q'_n (u) &=& i^{n + 1} Q'_n (iu), \\ 
 q'_0 (u) &=&  - \frac{1}{{u^2  + 1}}. 
\end{eqnarray}
Hence that for  $n=0$ the solutions is  
\begin{subequations}\begin{eqnarray}
e^{\psi} &=& \frac { 2 }{ (1+a)e^{c_0 \cot^{-1} u } +(1-a)e^{-c_0 \cot^{-1} u}   }, \\
\Lambda &=& - \frac{c_0^2}{2} \ln \left[ \frac{u^2+1}{u^2+v^2} 
\right ], \\
A &=& kbc_0 \left( {v + 1} \right) .
\end{eqnarray}\end{subequations}

This solution is the  magnetized version of the  Bonnor-Sackfield vacuum solution or zero order Morgan-Morgan solution \cite{BS,GG1}.  The solution is regular on the axis of symmetry but is not asymptotically flat and  represents the exterior gravitational field 
of finite disklike source   in the presence of a magnetic field
spread throughout the space (infinite).

For the terms  $n=0$ and  $n=2$ of series  (\ref{EQ:serieobla})  we have 
\begin{equation}
\hat  \psi   = - c_0 \cot^{ - 1} (u) - \frac{1}{4}c_2  
\left( 3v^2  - 1 \right) \left[ (3u^2  + 1) \cot^{ - 1} (u) - 3u \right].
\end{equation}

For the second term of the summation we have   
\begin{eqnarray}
 A  &=& -kb\frac{{c_2 }}{5}\left( {u^2  + 1} \right)q'_2 (u)\left[ {vP_2 (v) - P_1 (v)} \right] \nonumber \\ 
   &=&   \frac{1}{2}kbc_2 \left( {u^2  + 1} \right)v\left( {1 - v^2 } \right)q'_2 (u) \nonumber \\ 
   &=& \frac{1}{2}kbc_2 v\left( {1 - v^2 } \right)\left[ {3u\left( {u^2  + 1} \right) \cot^{ - 1} (u) - 3u^2  - 2} \right].
\end{eqnarray}

Finally, the solutions is 

\begin{subequations}\begin{eqnarray}
A &=& kbc_0 \left( {v + 1} \right) + \frac{1}{2}kbc_2 v\left( {1 - v^2 } \right)\left[ {3u\left( {u^2  + 1} \right)\cot^{ - 1} (u) - 3u^2  - 2} \right],   \nonumber  \label{eq:morgan1} \\
& & \\
\Lambda &=& 
-\frac{9}{16 k^2}c_2^2 \rho^2 (9u^2v^2-u^2+v^2-1) \cot^{ - 1}(u)^2 
\nonumber \\
&& +\frac{3}{8}c_2u(1-v^2)(27c_2u^2v^2-3c_2u^2+21c_2v^2-5c_2+8c_0)
\cot^{ - 1}(u)  \nonumber \\
&& + \frac{1}{2} (c_2-c_0)^2  \ln \left ( \frac{u^2+v^2}{1+u^2} \right )
\nonumber \\
&& - \frac{3}{16}c_2(1-v^2)(12c_2v^2+27c_2u^2v^2 -3c_2u^2 - 4c_2+16c_0),
\label{eq:morgan2}
\end{eqnarray}\end{subequations}
and again $\psi$ is given by  (\ref{eq:f}). In the absence of magnetic field $b=0$ and $c_0=c_2$,  this solution corresponds  the first order  Morgan-Morgan  vacuum  solution. Therefore
the case  $b\neq0$ corresponds  to a  magnetized first order Morgan-Morgan  solution.


\subsection{Solutions in generalized spheroidal coordinates}

We can reunite the three general solutions presented  above in a single family by
considering   generalized spheroidal coordinates ($\xi$,$\eta$) which are a generalization of the spherical coordinates, prolate and oblate.
The   asymptotically flat general solution of Laplace's equation  (\ref{eq:laplace}) can be written in 
this coordinates   
\begin{equation}
\hat  \psi (\varepsilon ,\eta ) = - \sum\limits_{n = 0}^\infty  {\frac{{c_n }}{{\sigma ^{n + 1} }}Q_n (\varepsilon )P_n (\eta )} ,
\end{equation}
where  $c_n$ are constants and  $\varepsilon= \xi / \sigma$. 
The expressions corresponding to the metric functions and the magnetic
potential are  obtained in the same way as the previous cases, that is 
\begin{subequations}\begin{eqnarray}
A(\varepsilon ,\eta ) &=&  \int_{ - 1}^\eta  {A ,_\eta  d\eta }, \\
\Lambda (\varepsilon ,\eta ) &=& \int_{ - 1}^\eta  {\hat  \Lambda ,_\eta  d\eta }.
\end{eqnarray}\end{subequations}

So the magnetic potential $A$ is 
\begin{equation}
A = -kb\left( {\xi ^2  -\sigma ^2 } \right)\sum\limits_{n = 0}^\infty  {\frac{{c_n }}{{(n+1)\sigma ^{n + 1} }}Q'_n  (\varepsilon )
 \left[ \eta {P_n (\eta ) - P_{n - 1} (\eta )} \right]} ,
\end{equation}
and the metric function  $\Lambda$  is given by \cite{QUEVEDO,RAMOS}
\begin{equation}
\Lambda  = \sum\limits_{n,m = 0}^\infty  {\frac{{c_n c_m }}{{\sigma ^{n + m + 1} }}\Gamma ^{mn} } ,
\end{equation}
where $Q'_n (\xi )$ is the total derivate of $Q_n (\xi )$ with respect to  $\xi$ and 
\begin{eqnarray}
\Gamma ^{mn}  &=& \frac{1}{2}\ln \left[ {\frac{{\varepsilon ^2  - 1}}{{\varepsilon ^2  - \eta ^2 }}} \right] + \left( {k_n  + k_m  - 2k_n k_m } \right)\ln \left[ {\frac{{\varepsilon  + \eta }}{{\varepsilon  - 1}}} \right] \nonumber \\
&& + \left( {\varepsilon ^2  - 1} \right)\left[ {\varepsilon \left( {A_{n,m} Q'_n Q_m  + A_{m,n} Q'_m Q_n } \right) - C_{n,m} Q_n Q_m } \right] \nonumber \\
&&+ \left( {\varepsilon ^2  - 1} \right)\left[ {\left( {1 - k_n } \right)S_m  + k_n S_{m + 1}  - \frac{{k_n }}{{m + 1}}\left( {P_m  - \left( { - 1} \right)^m Q'_m } \right)} \right] \nonumber \\
&&+ \left( {\varepsilon ^2  - 1} \right)^2 \left[ {Q_m B_{m,n}  - Q'_m A_{m,n}  + \frac{1}{{n + 1}}A_{m,n} Q'_m Q'_n } \right],
\end{eqnarray}
with
\begin{equation}
k_l  = \left\{ \begin{array}{l}
 1{\text{ for $l$ even}} \\ 
 0{\text{ for $l$ odd}} 
 \end{array} \right.
\end{equation}
and where $A_{n,m}$, $B_{n,m}$, $C_{n,m}$ and  $S_n$ are
\begin{subequations}\begin{eqnarray}
 A_{n,m} & +& A_{m,n}  = P_n P_m  - \left( { - 1} \right)^{n + m} \int_{ - 1}^\eta  {P'_m P_n d\eta }  ,\\ 
 B_{n,m}  &=& B_{n,m}  = \int_{ - 1}^\eta  {P'_m P_n d\eta } , \\ 
 C_{n,m}  &=& C_{n,m}  = \int_{ - 1}^\eta  {\eta P'_m P_n d\eta }  ,\\ 
 S_n  &=& \sum\limits_{k = 0}^{(n - 2)/2} {\left[ {\frac{1}{{n - 2k}} + \frac{1}{{n - 2k - 1}}} \right]\left( {P_{n - 2k - 1}  + \left( { - 1} \right)^{n + 1} } \right)} Q'_{n - 2k - 1} \nonumber .\\
&&
\end{eqnarray}\end{subequations}

Finally, the function $\psi$ is given by equation (\ref{eq:f}), and with 
$b \neq 0$ the solutions are asymptotically flat for  $ n \geq 1 $,
i.e., if the coefficient $c_0$ is chosen equal to zero. So  in presence of magnetic field all discussed solutions 
are either singular, not asymptotically flat or massless.


\section{Relativistic thin disks} 

In order to obtain a solution of the Einstein-Maxwell equations (\ref{eq:einmax1}) - (\ref {eq:einmax2}) 
representing a thin disk at $z=0$ with current, we assume  that the components of the metric tensor and of the electromagnetic potential are continuous across the disk, but with first  derivatives
discontinuous on the  plane $z=0$, with  discontinuity functions 
\begin{eqnarray}
b_{ab} \ &=& g_{ab,z}|_{_{z = 0^+}} \ - \ g_{ab,z}|_{_{z = 0^-}}
 \ = \ 2 \ g_{ab,z}|_{_{z = 0^+}}, \\
 a_{a} \ &=& A_{a,z}|_{_{z = 0^+}} \ - \ A_{a,z}|_{_{z = 0^-}}
 \ = \ 2 \ A_{a,z}|_{_{z = 0^+}}.                 
\end{eqnarray}
By using the distributional approach \cite{PH,LICH,TAUB} or the junction
conditions on the extrinsic curvature of thin shells \cite{IS1,IS2}, the
Einstein-Maxwell equations yield an  energy-momentum tensor $T_{ab}=
T^{\mathrm{elm}}_{ab} +  T^{\mathrm {mat}}_{ ab} = 
T^{\mathrm{elm}}_{ab} +  Q_{ab} \ \delta (z) $, 
and a  current density 
$J_a =j_a  \delta (z)= -  e^{2 (\psi - \Lambda)} a_a  \delta (z)$, where $\delta
(z)$ is the  usual Dirac function with support on  the disk,  $T^{\mathrm {
elm}}_{ab}$ is the electromagnetic tensor defined in Eq. (\ref{eq:tab}),
 $j_a$
is the current density on the plane  $z=0$, and 
$$
Q^a_b = \frac{1}{2}\{b^{az}\delta^z_b - b^{zz}\delta^a_b +  g^{az}b^z_b -
g^{zz}b^a_b + b^c_c (g^{zz}\delta^a_b - g^{az}\delta^z_b)\}
$$
is the distributional energy-momentum tensor. The ``true''  surface
energy-momentum tensor (SEMT) of the  disk,  $S_{ab}$, and the ``true'' surface
current density,  $\mbox{\sl j}_a$, can be obtained through the
relations  	
\begin{subequations}\begin{eqnarray}
S_{ab} &=& \int T^{\mathrm {mat}}_{ab} \ ds_n \ = \ e^{  \Lambda - \psi} \ 
Q_{ab} \ ,   \\
\mbox{\sl j}_a  &=& \int J_{a}  \ ds_n \ = \ e^{ \Lambda -  \psi} \ j_a \ , 
\end{eqnarray}\end{subequations}
where $ds_n = \sqrt{g_{zz}} \ dz$ is the ``physical  measure'' of length in the
direction normal to the disk. For the metric (\ref{eq:met}), the nonzero components of  $S_a^b$ are
\begin{subequations}\begin{eqnarray}
&S^0_0 &= \ 2 e^{\psi - \Lambda} \left\{ \Lambda,_z - \ 2 \psi,_z  \right\} ,
 \label{eq:emt1}     		\\
&	&	\nonumber	\\
&S^1_1 &= \ 2 e^{\psi - \Lambda} \Lambda,_z , \label{eq:emt2}
\end{eqnarray}\label{eq:emt}\end{subequations}
and the only nonzero component of the current density  $\mbox{\sl j}_a$
in the  magnetostatic case is     
\begin{equation}
\mbox{\sl j}_{\varphi} = \ - \frac{1}{2 \pi} e^{\psi - \Lambda} 
A _{,z},  \label{eq:cormag} 
\end{equation}
where all the quantities are evaluated at $z = 0^+$.

In order to give physical significance  to the components of  the
energy-momentum tensor  $S_a^b$ and  the electric current density  $\mbox{\sl j}_a$  we project them onto the   orthonormal tetrad ${{\rm e}_{ (a)}}^b = \{ V^b , W^b , X^b,Y^b \}$, where
\begin{subequations}\begin{eqnarray}
V^a &=& e^{- \psi} \ ( 1, 0, 0, 0 ) ,	\\
	&	&	\nonumber	\\
W^a &=& \frac{e^\psi} {\rho} \ \ ( 0, 1, 0, 0 ) ,	\\
	&	&	\nonumber	\\
X^a &=& e^{\psi - \Lambda} ( 0, 0, 1, 0 ) ,	\\
	&	&	\nonumber	\\
Y^a &=& e^{\psi - \Lambda} ( 0, 0, 0, 1 ) .
\end{eqnarray}\label{eq:tetrad}\end{subequations}
In terms of this tetrad (or observer with four-velocity $V^a$)
the surface energy density $\epsilon$, the  
azimuthal pressure $p_\varphi$, and the azimuthal current density $\mbox{\sl j}$  of the disk are given by
\begin{equation}
\epsilon \ = \ - S^0_0 , \quad p_\varphi \ = \ S^1_1 , \quad  \mbox{\sl j}  = W^1 \mbox{\sl j}_1.
\label{eq:jtet}
\end{equation}


Finite thin disks can be
obtained  introducing oblate  spheroidal coordinates,  which are  naturally
adapted to a disk source. These
solutions, in the vacuum and static case, correspond to the Morgan and Morgan
solutions \cite{MM1}. In the case of the magnetized  Morgan-Morgan solutions, following the reference \cite{MM1},  for $n$ even 
the metric functions
are  continuous across the disk but its  first derivatives are discontinuous in the direction normal to the disk, which can be interpreted as a finite thin disk 
located at $z = 0$ and $0 \leq \rho \leq 1$.  However,
since for $n$ even the function $A$ is a odd polynomial of $v$,  
the opposite occurs with the magnetic potential, i.e,   it is   discontinuous across the disk but its first  derivatives are continuous  in the direction normal to the disk, so that  electric current density  also is  zero on the disk  and in consequence the source of the
magnetic field is non planar but of a different origin such as a remnants or fossil magnetic field \cite{FOSSIL}, or can come from
external sources, such as the presence of a nearby magnetars or 
neutron stars. Indeed on the disk the Maxwell equations  $\partial_ b \bar F^{ab} = 4 \pi \bar J^a$, where `bar' denotes multiplication by $\sqrt{-g}$,   are given by 
\begin{equation}
-4 \pi \bar j_\varphi \delta(z) =  \partial _z g^{zz} \bar A_{,z} 
+  \partial _{\rho} g^{\rho \rho} \bar A_{,\rho},
\end{equation}
where $\delta(z)$ is the usual Dirac Function with support on the disk. Integrating through 
the disk we obtain
\begin{eqnarray}
-4 \pi \bar j_\varphi  & = &  \int _{z=0_-}^{z=0_+}  \partial _z g^{zz} \bar A_{,z} dz
+\int _{z=0_-}^{z=0_+} \partial _{\rho} g^{\rho \rho} \bar A_{,\rho} dz 
\nonumber  \\
&=&  \left. g^{zz} \bar A_{,z} \right |_{z=0_-}^{z=0_+} + \partial _{\rho} g^{\rho \rho} \partial_{\rho}  \int _{z=0_-}^{z=0_+} \bar A dz  \nonumber \\
&=& 0, 
\end{eqnarray}
where the first  term on the right-hand side vanishes due to the  continuity of the metric and  $A_{,z}$ and the second term
from discontinuity of $A$, or  in other words  as  $A$ is an odd  polynomial function  of $v$  its integral through 
the disk is  even and hence continuous, so that to the  evaluate 
at the limits of integration expression vanishes. 
Thus we can interpret  this solutions
as  the exterior gravitational field of  a finite dislike source  
immersed in a magnetic field.
For completeness, we analysis   the electrostatic case, i.e., for $\Phi$ real. The electric potential $\phi$ is given by 
\begin{equation}
\phi =  \frac{p (e^{-\hat  \psi } - e^{\hat  \psi })}{ (1+a )e^{ - \hat  \psi }  
+ (1-a) e^{\hat  \psi } }, \label{eq:ep}
\end{equation}
where  $a=\sqrt{1+p^2}$, being   $p$  the parameter that controls the
electric field, and the only non-zero component of the current density 
$\mbox{\sl j}_a$ on the plane $z=0$ is 
\begin{equation}
\mbox{\sl j}_t = \ - \frac{1}{2 \pi} e^{\psi - \Lambda} \phi _{,z}, 
\label{eq:corelec}  
\end{equation}
whereas in terms of the tetrad (\ref{eq:tetrad}) the  electric charge density 
$\sigma$   is given  by
\begin{equation}
\sigma = -V^0 \mbox{\sl j}_0.
\label{eq:sigmatet}
\end{equation} 

We see that just as  metric the electric potential $\phi$  is  a even function of $v$ and in consequence the current density in the disk is nonzero. 
Therefore, this solutions can be interpreted as the gravitational field of  a finite charged disk. 

\begin{figure*}
$$
\begin{array}{ccc}
\epsilon    &  p_\varphi  &  \sigma  \\
\epsfig{width=2.0in,file=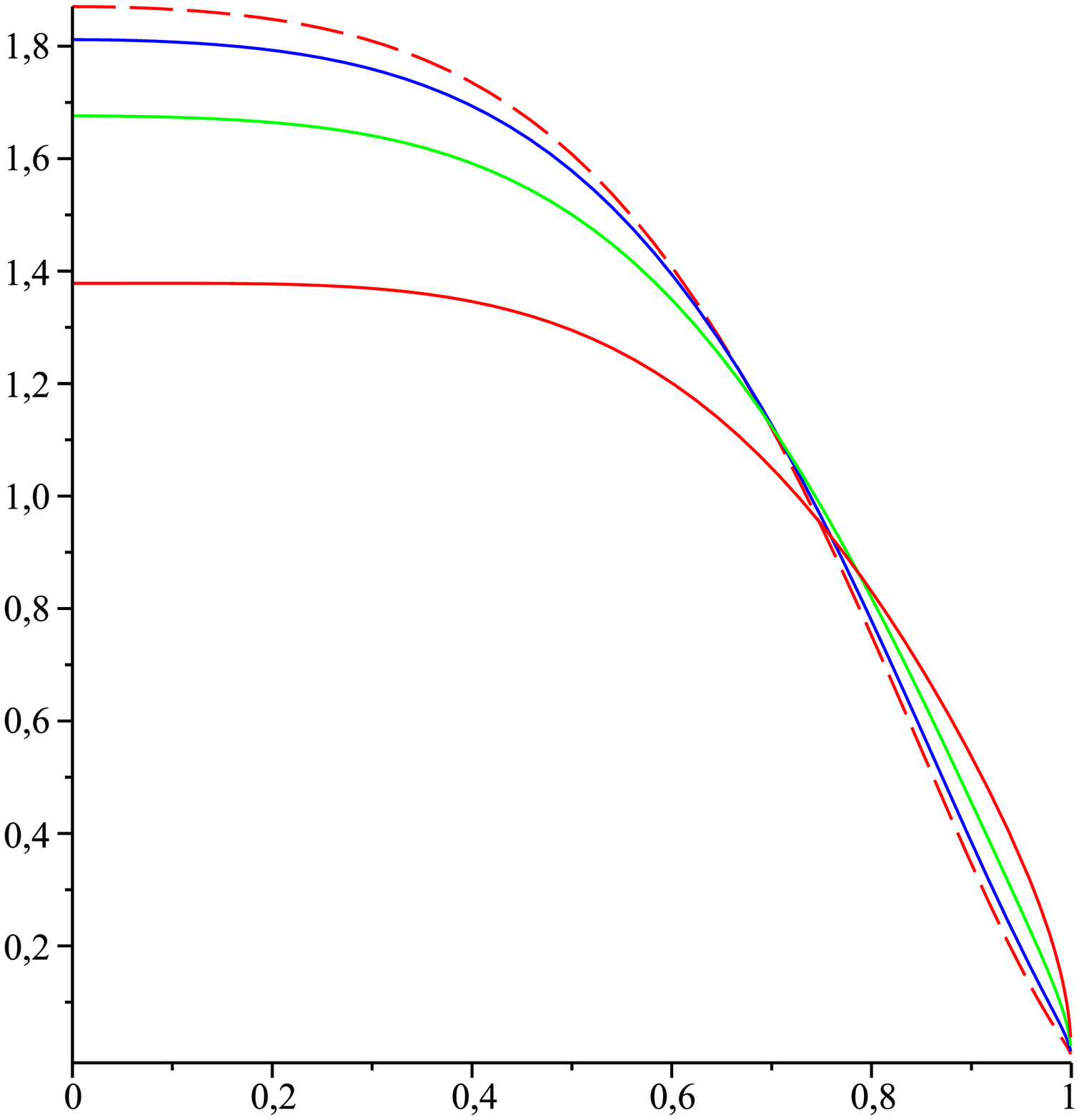} &
\epsfig{width=2.0in,file=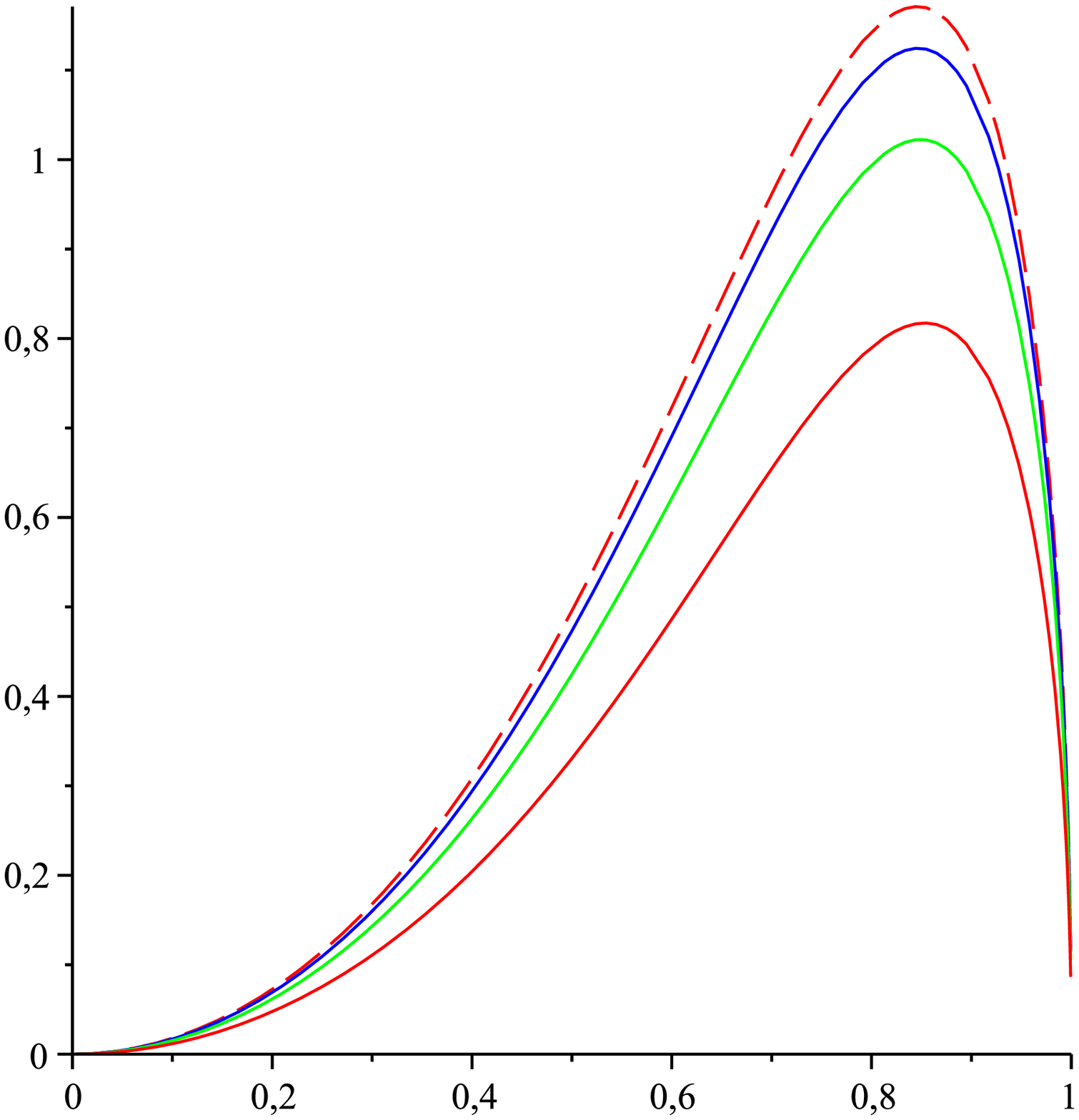}   &
\epsfig{width=2.0in,file=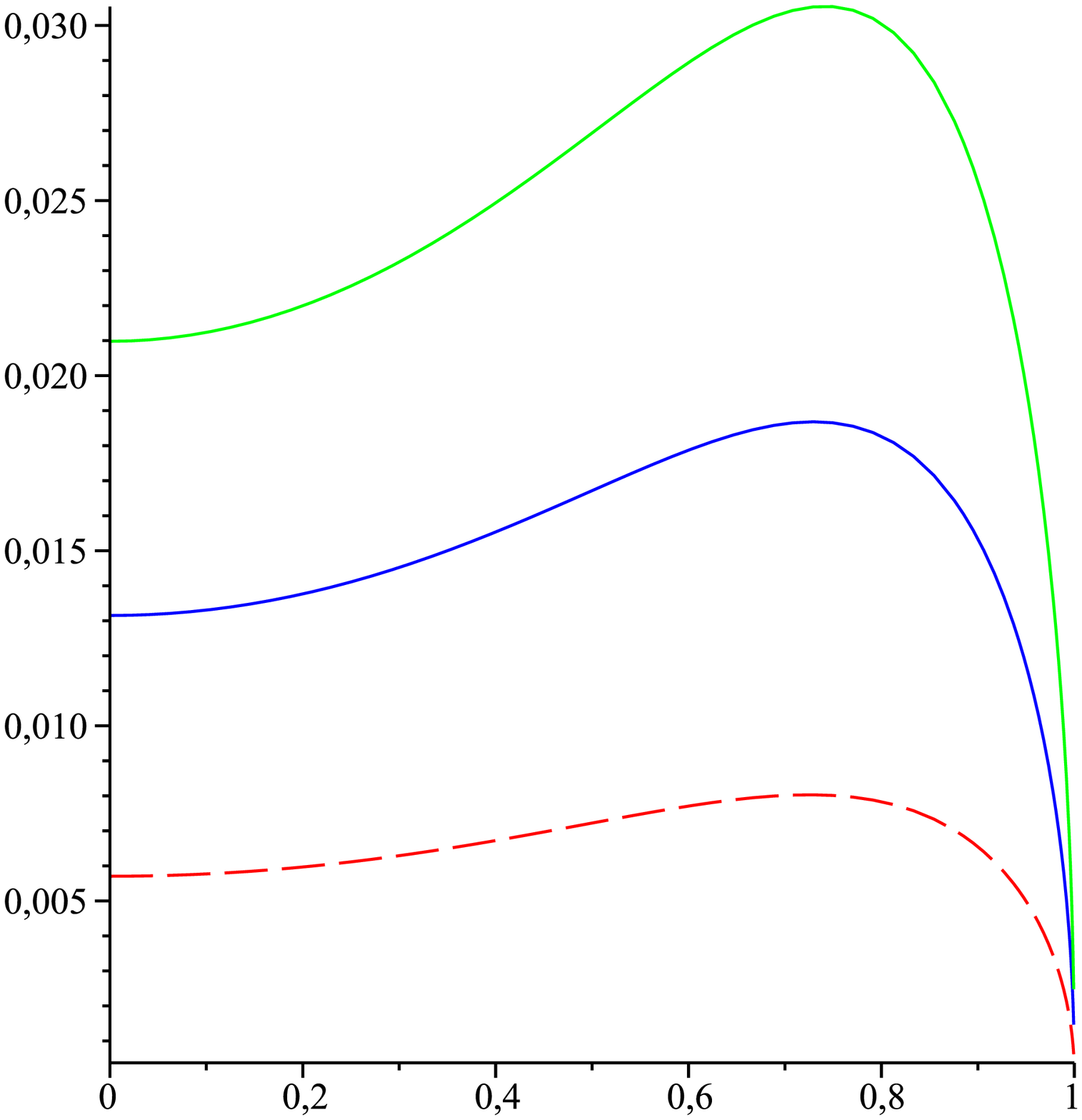}  \\
\rho &  \rho  & \rho  \\
(a)     &   (b) & (c) \\
\end{array}
$$	
\caption{$(a)$  The surface energy density $\epsilon$  and $(b)$ the azimuthal pressure $p_\varphi$  for magnetized first order Morgan-Morgan 
finite disks  with $c_0=c_2=0.4$ and  for  values of magnetic field parameter  $b= 0$ (dashed curves), $0.5$,  $1$, and $2$ (bottom curves), as  functions of $\rho$. $(c)$ The surface electric charge density $\sigma$ (electrostatic case) for $p= 0.2$ (dashed curve), $0.5$, and  $1$ (top curve) and the same values of $c_0$ and $c_2$. }
\label{fig:enerprecar}
\end{figure*}

In order to study the behavior of the main  physical quantities associated with these disks  we perform a graphical analysis of them
for magnetized first order Morgan-Morgan  finite disks. 
In Fig. \ref{fig:enerprecar} we show  the energy density 
$\epsilon$  and the azimuthal pressure $p_\varphi$  with $c_0=c_2=0.4$ and  for  values of magnetic field parameter  $b= 0$ (dashed curves), $0.5$,  $1$, and $2$ (bottom curves), as  functions of $\rho$. We  see  that the energy density presents a
maximum at $\rho = 0$ and then decreases  with  $\rho$. We also see that  the
presence of magnetic  field decreases the energy density  at the central
region of the disk and  later increases it.  We can observe that the pressure increases rapidly as one moves away from the disk center, reaches a maximum
and  later  decreases. We also observe that the magnetic field
decreases the pressure everywhere on the disk.
The graph also show that the disk's surface energy density is always positive in concordance with  the weak energy condition, as well as     the stress in azimuthal  direction (pressure).
The strong energy condition, $ \epsilon + p_\varphi >0 $,  is also satisfied. These properties characterize a fluid made of matter
with the usual gravitational attractive property.
In addition, in Fig. \ref{fig:enerprecar}$(c)$ we show, in the electrostatic case,  the surface electric charge density $\sigma$ for  values of electric  field parameter  $p= 0.2$ (dashed curve), $0.5$, and  $1$ (top curve) and the same values of $c_0$ and $c_2$.
We  observe a greater concentration of electric charge near the rim of the disk and that this  increases as we increase the electric field.


Exact solutions which represent the field of a  disk also
can be obtained using  the well known ``displace,
cut and reflect'' method  that was first
used by Kuzmin \cite{KUZMIN} and Toomre \cite{TOOMRE} to constructed Newtonian models of disks, and later extended to general relativity 
\cite{BLK,BLP,BL,GL2}. Given a solution of the
Einstein-Maxwell equation, this procedure is mathematically equivalent to apply the transformation $z \rightarrow |z| + z_0$,
with $z_0$ constant, on that solution.  However, this disks are essentially  of infinite extension and the field  not correspond exactly to the metric
with which we started.  These  solutions can be called in our
case  magnetized Weyl type solutions.  The reflection symmetry with respect
to the plane of disk  implies  that in this case  $A$  is continuous across the disk and its first derivative discontinuous in the direction normal 
to the disk, so that  electric current density   is  nonzero on the disk  and in consequence the source of the magnetic field is planar. 

\begin{figure*}
$$
\begin{array}{ccc}
\epsilon    &  p_\varphi &   -\mbox{\sl j} \\
\epsfig{width=2.0in,file=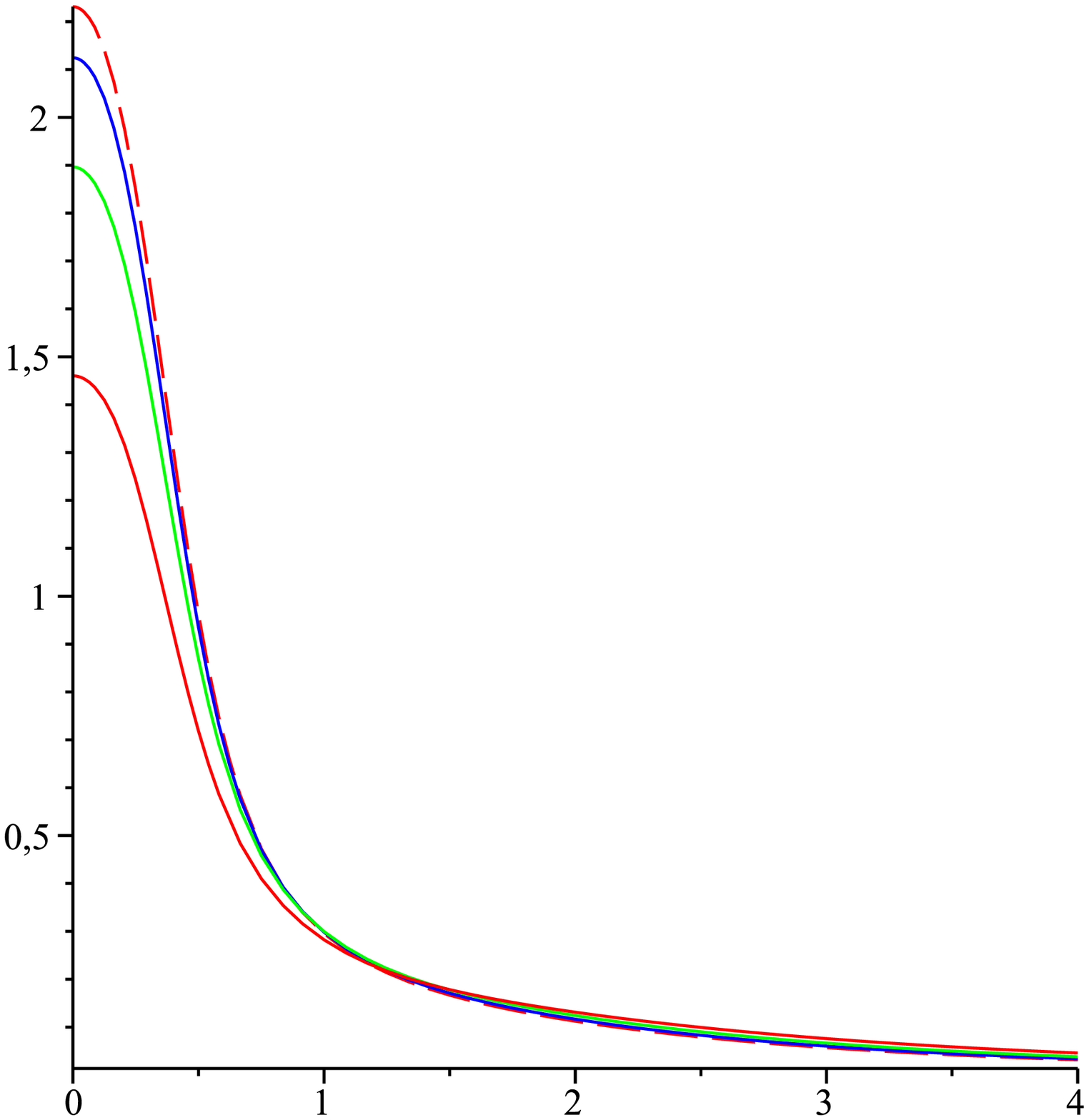} &
\epsfig{width=2.0in,file=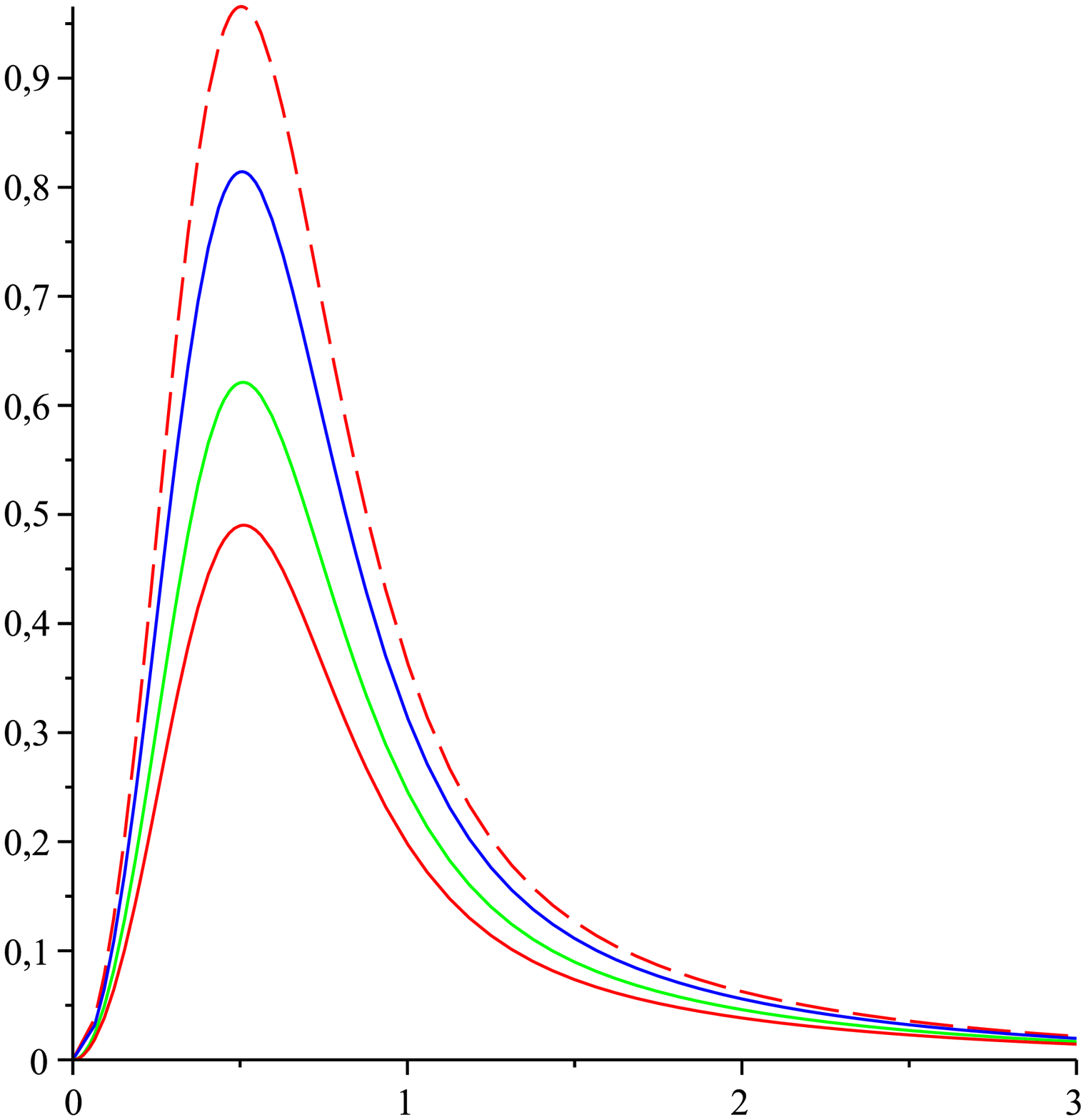} &
\epsfig{width=2.0in,file=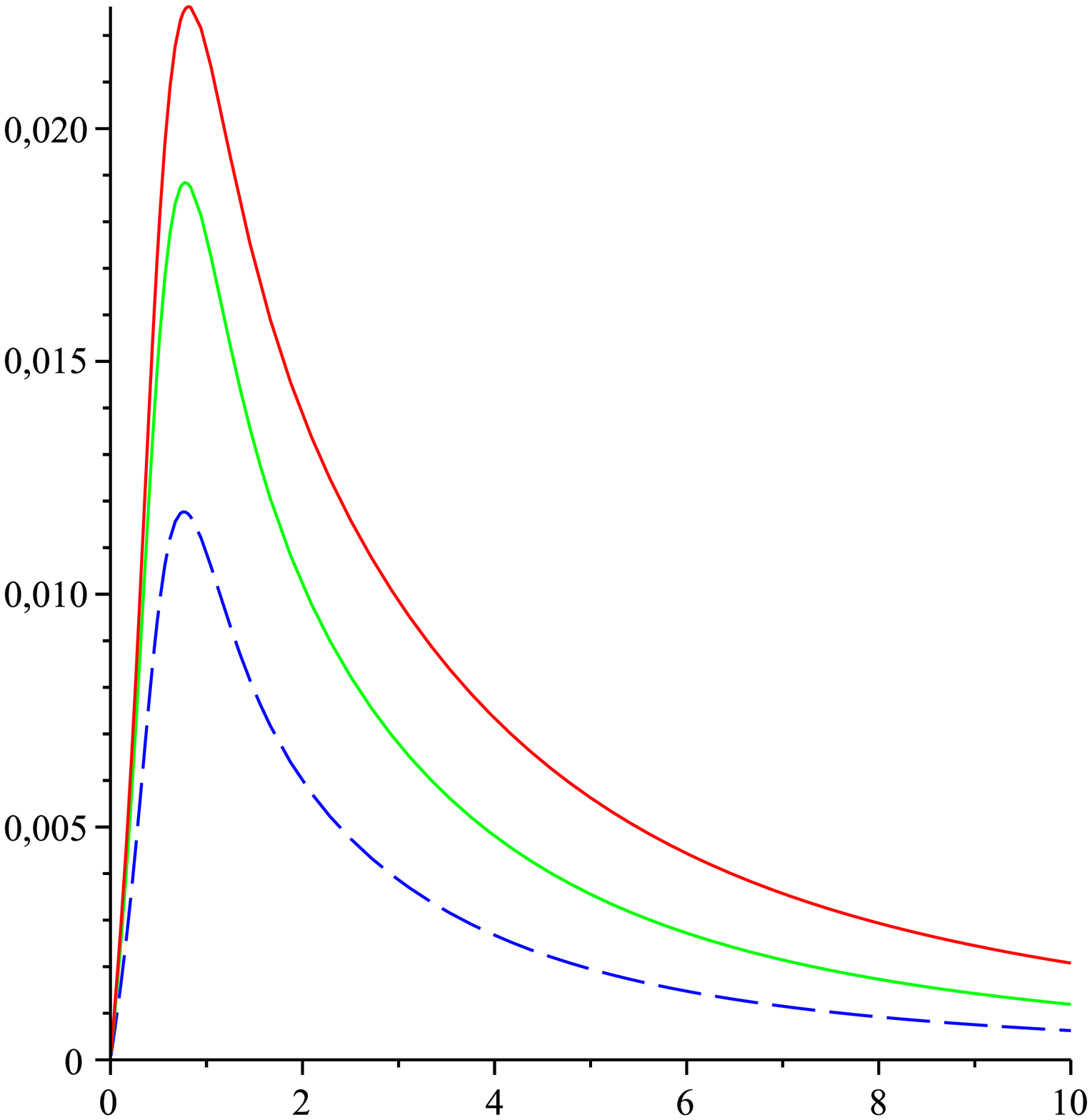}  \\
 \rho &  \rho  & \rho  \\
(a)     &   (b)  & (c) 
 \end{array}
$$	
\caption{$(a)$  The surface energy density $\epsilon$  and $(b)$ the azimuthal pressure $p_\varphi$  for  magnetized weyl  type  infinite disks 
in spherical  coordinates and second order with $z_0=1$,  $c_0=1$, $c_2=0.5$ and  for  values of magnetic field parameter  $b= 0$ (dashed curves), $0.5$,  $1$, and $2$ (bottom curves), as  functions of $\rho$. $(c)$ The azimuthal current density $\mbox{\sl j}$ for $b= 0.5$ (dashed curve), $1$, and  $2$ (top curve) and the same values of $z_0$, $c_0$ and $c_2$. }
\label{fig:enprj-esfe}
\end{figure*}

\begin{figure*}
$$
\begin{array}{ccc}
\epsilon    &  p_\varphi & -\mbox{\sl j}   \\
\epsfig{width=2.0in,file=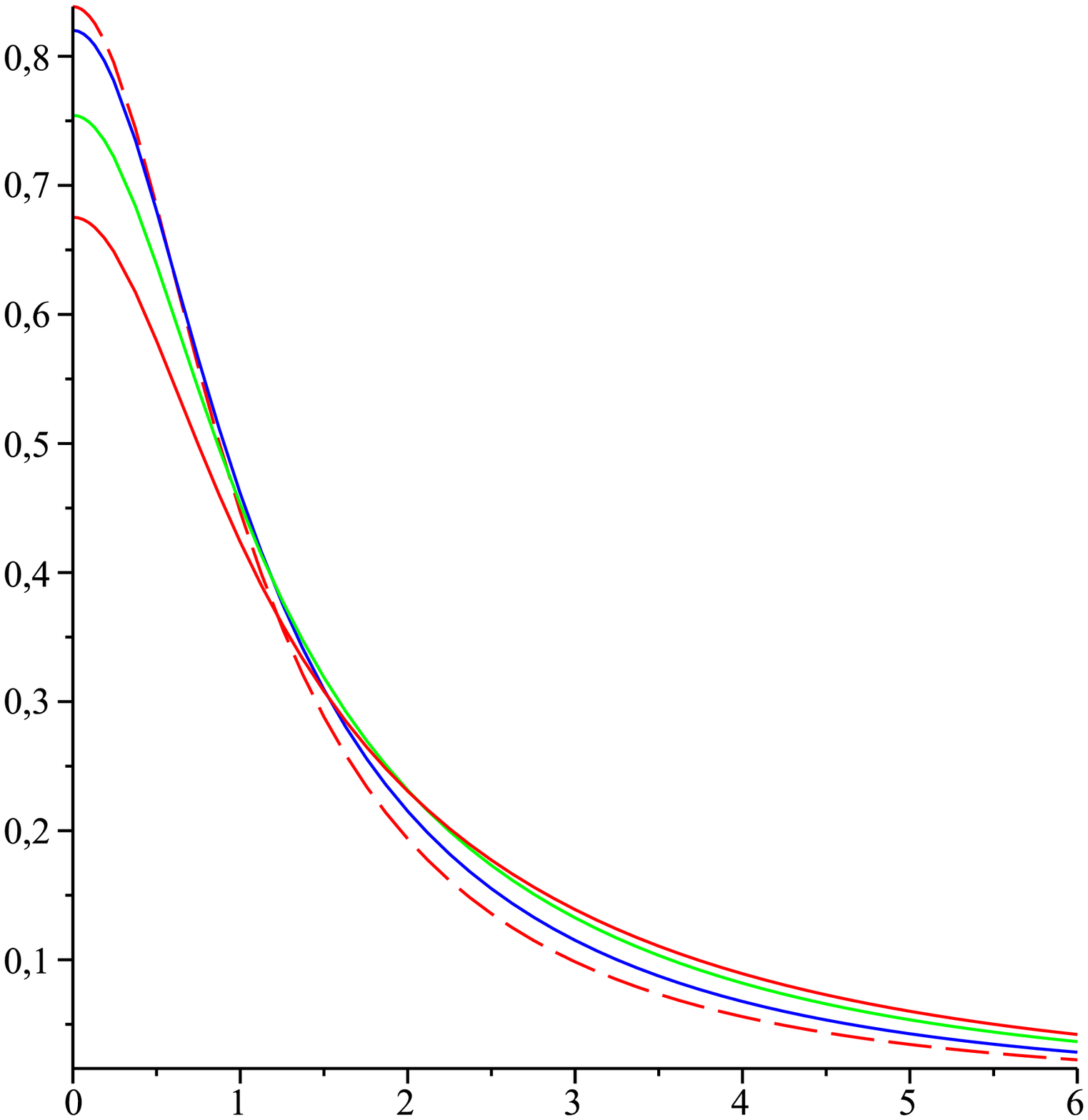} &
\epsfig{width=2.0in,file=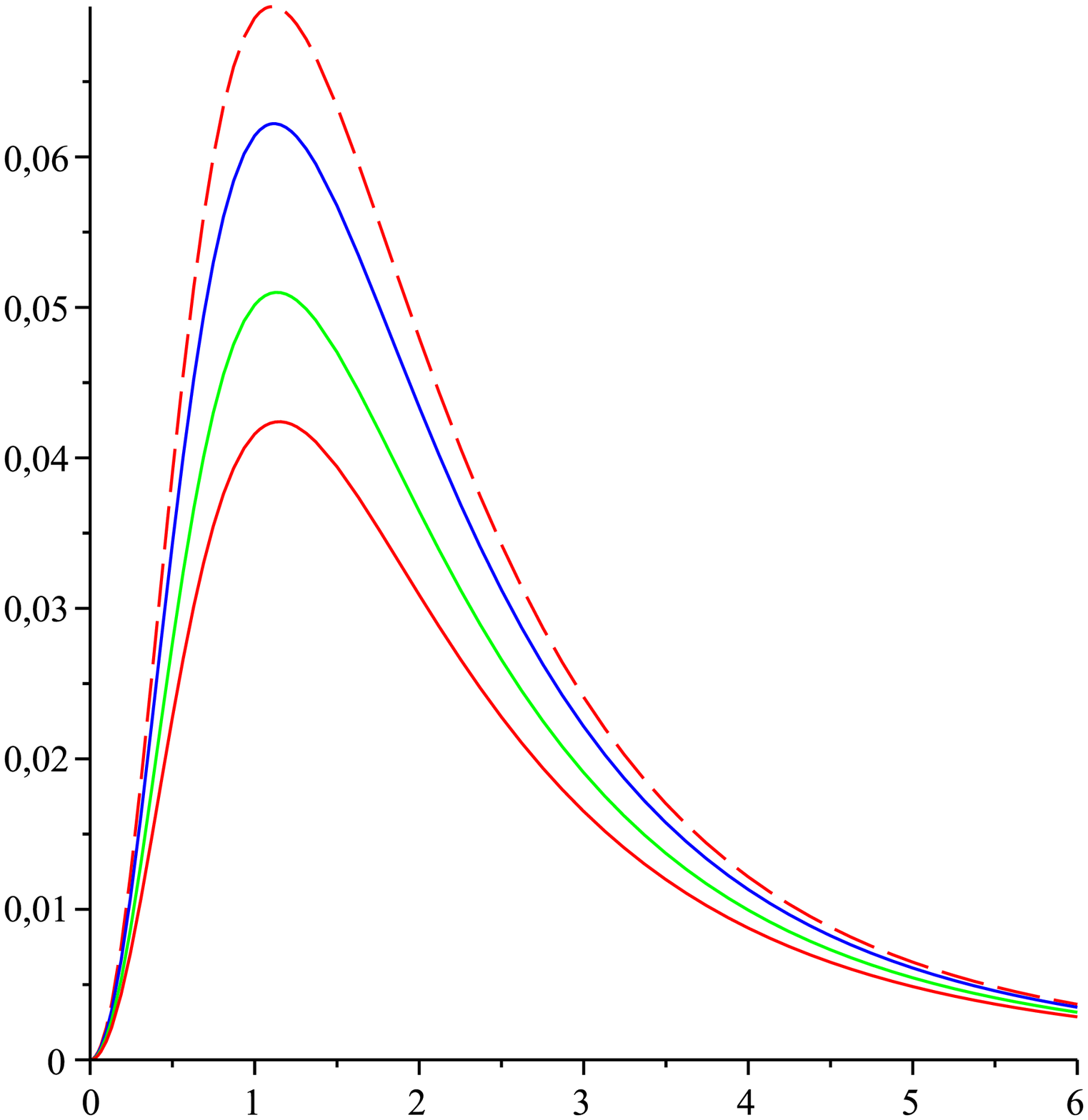} &
\epsfig{width=2.0in,file=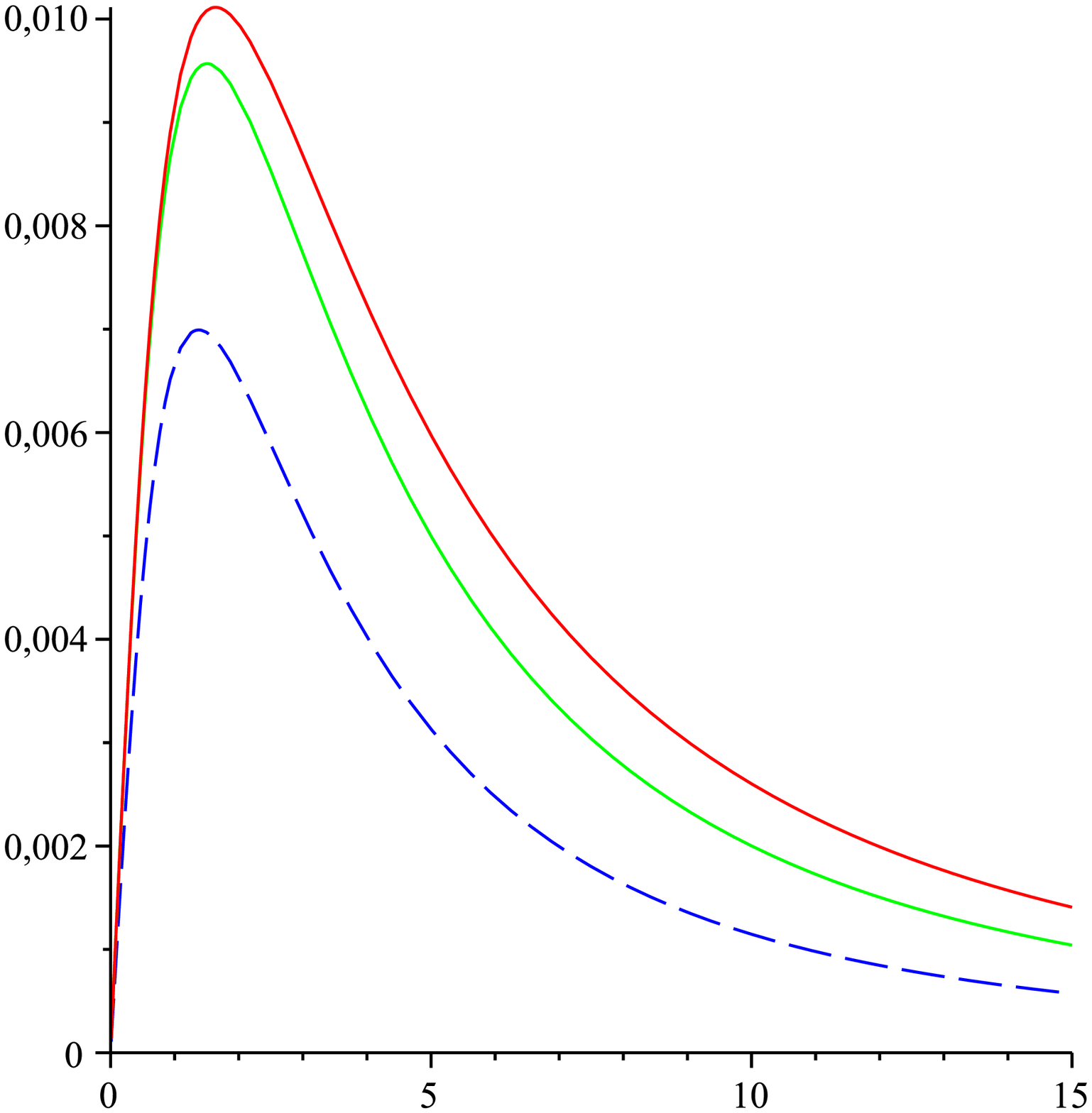} \\
 \rho &  \rho  & \rho  \\
(a)     &   (b) & (c) 
 \end{array}
$$	
\caption{$(a)$ The surface energy density $\epsilon$  and $(b)$ the azimuthal pressure $p_\varphi$  for magnetized Erez-Rosen type infinite disks 
with $z_0=2$,  $c_0=c_2=1$ and  for  values of magnetic field parameter 
$b= 0$ (dashed curves), $1$,  $2$, and $3$ (bottom curves), as  functions of $\rho$. $(c)$ The azimuthal current density $\mbox{\sl j}$ for $b= 1$ (dashed curve), $2$, and  $3$ (top curve) and the same values of $z_0$, $c_0$ and $c_2$. }
\label{fig:enprj-erez}
\end{figure*}

\begin{figure*}
$$
\begin{array}{ccc}
\epsilon    &  p_\varphi &  -\mbox{\sl j}  \\
\epsfig{width=2.0in,file=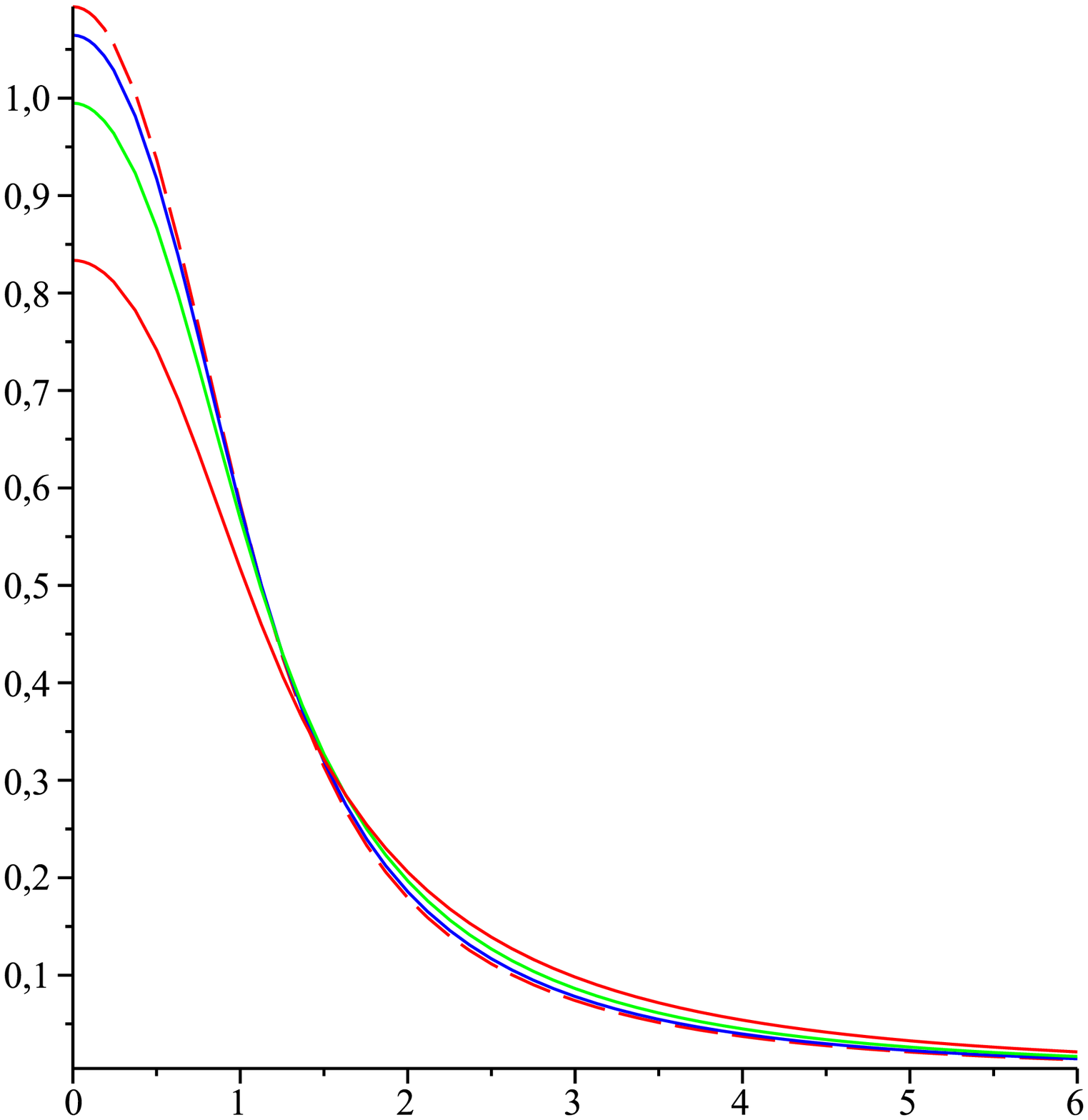} &
\epsfig{width=2.0in,file=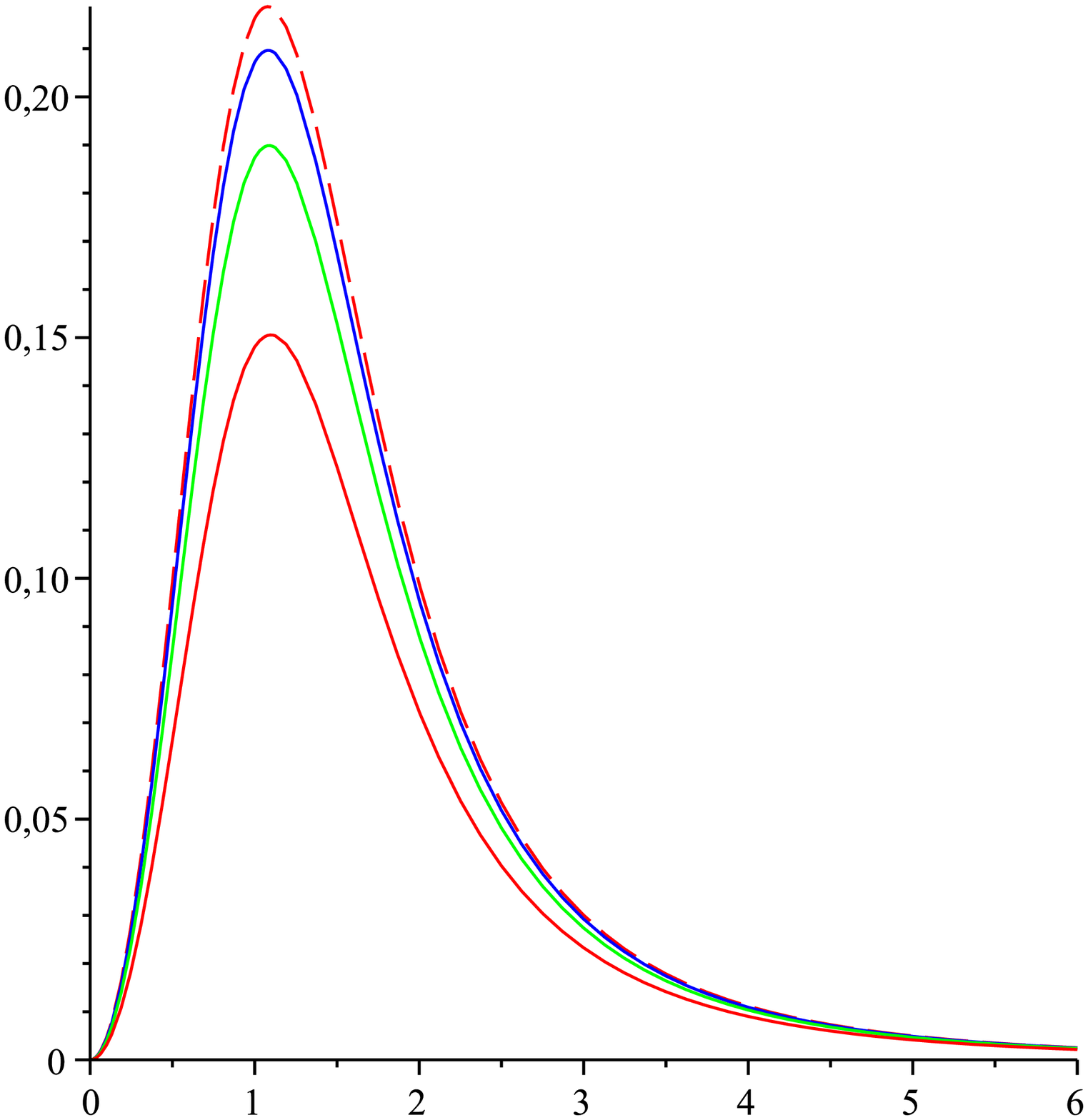} & 
\epsfig{width=2.0in,file=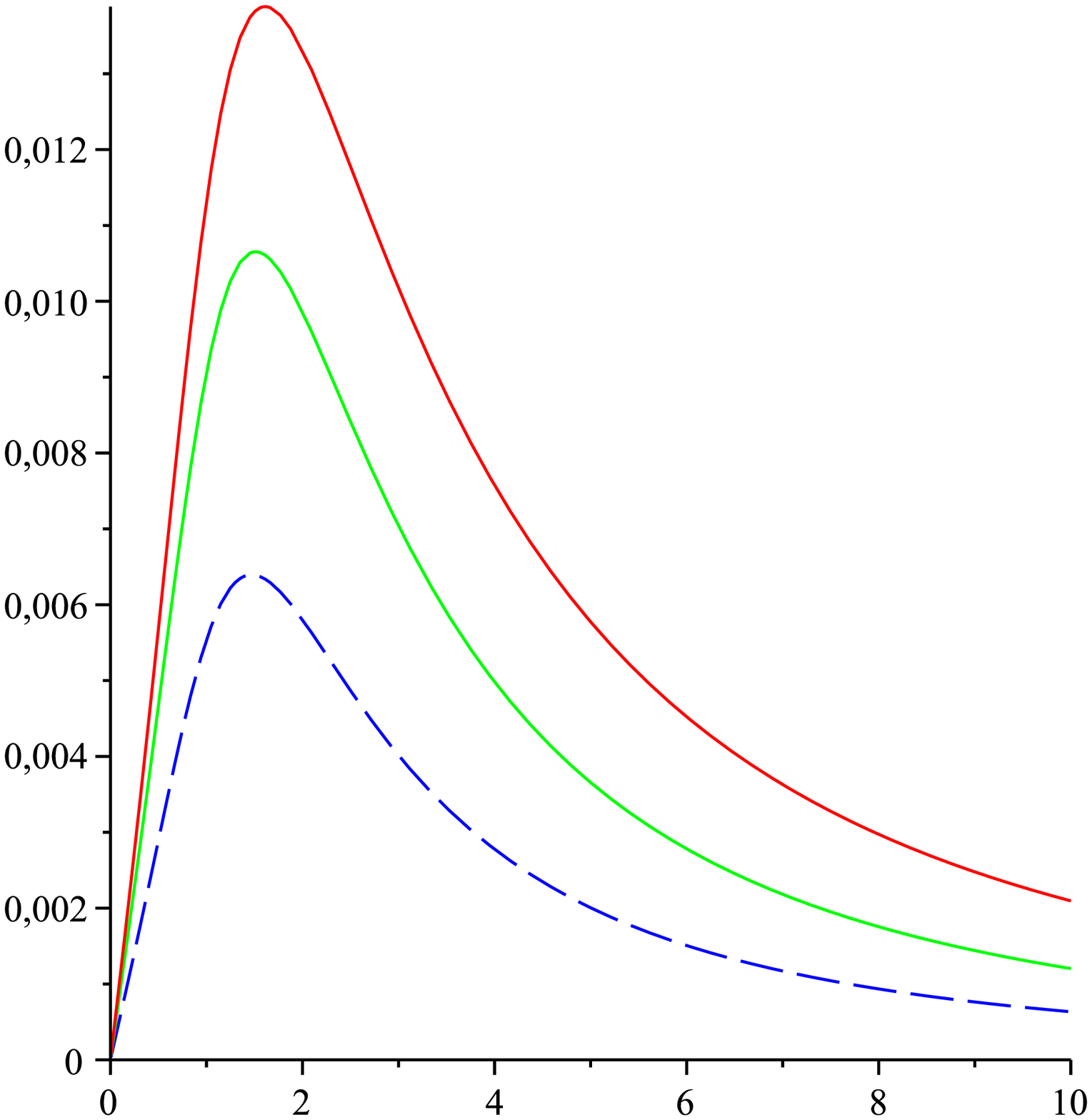}  \\
 \rho &  \rho  & \rho \\
(a)     &   (b) & (c) 
 \end{array}
$$	
\caption{$(a)$  The surface energy density $\epsilon$  and $(b)$ the azimuthal pressure $p_\varphi$  for magnetized first order  Morgan-Morgan type infinite disks with $z_0=c_0=c_2=1$ and  for  values of magnetic field parameter $b= 0$ (dashed curves), $0.5$,  $1$, and $2$ (bottom curves), as  functions of $\rho$. $(c)$ The azimuthal current density $\mbox{\sl j} $ for $b=0.5$ (dashed curve), $1$, and  $2$ (top curve) and the same values of $z_0$, $c_0$ and $c_2$. }
\label{fig:enprj-morgan}
\end{figure*}

In Figs. \ref{fig:enprj-esfe} - \ref{fig:enprj-morgan} we show, as  functions of 
$\rho$,   the energy density 
$\epsilon$, the azimuthal pressure $p_\varphi$ and  surface  azimuthal electric current density $\mbox{\sl j}$  for  magnetized Weyl type disks of  infinite extension corresponding to the lineal combination of the first and second terms of the series in the three families of solutions considered.
Since the surface energy density decreases
rapidly one can to define a cut off radius  and, in principle,  to consider these disks as finite. Anyway all the physical quantities present a similar behavior to the previous case.


\section{Motion of charged particles around magnetized Weyl fields}

The  relativistic Lagrangian for a  test particle in presence of a
gravitational and magnetic field is given by 
\begin{equation}
{\cal L} = \frac 12 g_{ab} \dot x ^a \dot x^b + \tilde e A \dot 
\varphi,
\end{equation}
where  $ \tilde  e = e / m$ is the specific charge of the particle and
the overdot denotes derivate with respect to the proper time $s$. For
magnetized Weyl fields we have two constants of motion 
\begin{eqnarray}
E &= &  - p_t / m= - g_{tt} \dot t, \\
L  & =&  p_{\varphi} / m =  g_{\varphi \varphi } \dot \varphi + \tilde  e A ,  
\end{eqnarray}
where $E$ represents the relativistic specific energy and $L$ the specific angular momentum.

For circular orbits and from symmetry of the field the equation for the electrogeodesic motion of the particle is given by
\begin{equation}
\frac 12   g_{ab,\rho}u^a u^b = - \tilde e  F_{\rho a} u^a_,
\label{eq:geo}
\end{equation}
where $u^a$ is the 4-velocity of particles with respect a the coordinates
frame.  For  equatorial circular orbits 
$u^a$ has components $u^a = u^0(1,\omega, 0,0 )$ where  $\omega= u^1/u^0=\frac{d \varphi}{d t}$ is  the 
angular velocity of the test particles, and  in the case magnetostatic 
the equation of motion  reads
\begin{equation}
 \frac 12 u^0 (g_{\varphi \varphi,\rho } \omega ^2 + g_{t t,\rho } )= - \tilde e A_{,\rho} \omega, \label{motion}
\end{equation}
where $u^0$ obtains normalizing  $u^a$, that is requiring  $g_{ab}u^au^b=-1$. Thus with 
\begin{equation}
 (u^0)^2 = - \frac {1}{g_{\varphi \varphi} \omega^2 + g_{tt}} \label{eq:u0}
\end{equation}
the equation of the  electrogeodesic takes the form
\begin{equation}
\bar A (\omega ^2)^2 + \bar B \omega ^2 + \bar C =0 ,
\end{equation}
where 
\begin{subequations}\begin{eqnarray}
\bar A & = & g_{\varphi \varphi, \rho}^2  + 4 \tilde e ^2  g_{ \varphi \varphi} A_{, \rho}^2  \nonumber  \\
& = & 4 \rho ^2 e^{- 2 \psi} [(1 -  \rho \psi_{, \rho})^2 e^{- 2 \psi} + \tilde e^2 A_{, \rho}^2   ],  \\
 \bar B & = & 2 g_{t t, \rho}  g_{\varphi \varphi, \rho} +  4 \tilde e ^2 g_{ t t}  A_{, \rho}^2   \nonumber \\
&=&  -4  [  2 \rho \psi_{, \rho} (1 -  \rho \psi_{, \rho}) + \tilde e^2 A_{, \rho}^2 e^{2 \psi} ],  \\  
\bar C & =&  g_{t t, \rho}^2 = 4  \psi_{, \rho} ^2  e^{ 4 \psi}.
\end{eqnarray}\end{subequations}

Therefore, the angular velocity $\omega $  is given by
\begin{equation}
\omega ^2 = \frac { - \bar B \pm \sqrt { D }}{ 2 \bar A  }, \label{eq:omega}
\end{equation}
where
\begin{equation}
D= \bar B^2 - 4 \bar A \bar C = 16 \tilde e ^2  A_{, \rho}^2 e^{2 \psi} [ 4 \rho \psi_{, \rho} (1 - 2 \rho 
\psi_{, \rho}) + \tilde e^2 A_{, \rho}^2 e^{2 \psi} ] \geq 0,
\end{equation}
and positive sign corresponds to the direct orbits (or co-rotating with
$L>0$) and the negative sign  to the retrograde orbits (or counter-rotating with $L<0$).  From (\ref{eq:u0}) and (\ref{eq:omega}) we find 
\begin{eqnarray}
E &= &  - g_{tt} u^0, \\
L  & =&  g_{\varphi \varphi } \omega u^0 + \tilde  e A . \label{eq:L}
\end{eqnarray}

With respect to the  orthonormal tetrad (\ref{eq:tetrad}) the 3-velocity
has components
\begin{equation}
 v^{ (i)}=  \frac { {{\rm e}^{ (i)}}_a u^a } { {{\rm e}^{(0)}}_b u^b } .              
\end{equation}
For equatorial circular  orbits  the 
only nonvanishing velocity components is given by 
\begin{equation}
 (v^{ (\varphi)})^2 = v_c^2= - \frac{ g_{\varphi \varphi} }{ g_{tt} } \omega ^2 = \rho ^2 {\rm e}^{-4\psi} 
\omega ^2,  \label{eq:vc2}
\end{equation}
which is the circular velocity of the particle as seen by an observer at infinity. In fact  when $\rho  \rightarrow \infty$, $\psi
\rightarrow 0$ and $v_c^2 = \rho ^2 \omega ^ 2 = \rho \frac{d \phi}{d \rho}$, where $\phi$ is the Newtonian gravitational potential,
 and  $v_c $  is the circular velocity, which represents  the velocity of
a test particle in a circular orbit at radius $\rho$. 
For neutral particles $\tilde e = 0$ we have
\begin{equation}
v_c^2 = \frac {\rho \psi_{, \rho}}{1 - \rho\psi_{, \rho} }, 
\label{eq:v2neutras}
\end{equation}
with
\begin{equation}
\psi_{,\rho}=  \hat \psi_{,\rho}  \left[ \frac { (1+a){\rm e}^{- \hat \psi} - (1-a) {\rm e}^{\hat \psi } }
{ (1+a){\rm e}^{-\hat \psi} + (1-a) {\rm e}^{\hat \psi } }  \right ].
\end{equation}
In terms of the circular velocity $v_c$ the specific angular momentum can be written as 
\begin{equation}
L =    \frac{ \rho e ^{- \psi} v_c}{ \sqrt { 1 - v_c^2} } + \tilde e A .
\end{equation}

To analyze the stability of circular orbits on the equatorial plane we
use  an extension of Rayleigh  criteria of stability of a fluid at rest
in a gravitational field  \cite{RAYL,FLU,LETSTAB}. The motion equation (\ref{motion}) can be cast as a balance equation
\begin{equation}
\frac {  g^{\rho \rho}  g_{tt,\rho} E^2 } {2 g_{ tt}^2 }
+ \tilde e g^{\rho \rho} A_{,\rho} \omega u^0 
 = - \frac { g^{\rho \rho}  g_{\varphi \varphi,\rho} (L - \tilde e  A)^2 } { 2 g_{\varphi \varphi}^2 }  \label{eq:equilibrio}
\end{equation}
where the term first on the left-hand side represents the gravitational force $F_g$, the term second  the the Lorentz force $F_L$,  and the term on the right-hand side  the centrifugal force $F_c(\rho)=F(\rho,L(\rho))$  acting on the particle. So  we have a balance between the total force $F(\rho) =F_g + F_L$ and the centrifugal force.   We now
consider the particle to be initially in a circular orbit with 
radius $\rho=\rho_0$ and we  slightly displace it to a higher orbit $\rho>\rho_0$. The angular momentum of particle  remains equal to its initial value  $L_0=L(\rho_0)$ which implies that the centrifugal force   in its new position is $ F_c(\rho,L_0)$. In order that the particle returns to it initial position must be met that  $F(\rho)>F_c(\rho,L_0)$, but 
according to the balance equation (\ref{eq:equilibrio}) $F(\rho)=F_c(\rho,L)$ so that
$F_c(\rho,L) > F_c(\rho,L_0)$,  and  hence $(L - \tilde e A)^2 > (L_0 - \tilde e A)^2 $. Using the expresion for $L$ (\ref{eq:L})  and defining 
la function $h=g_{\varphi \varphi } \omega u^0$,
follows that  $h(\rho)^2>h(\rho_0)^2$. Note that the quantity $h$ has the same form that the specific angular momentum in the vacuum. By doing a Taylor expansion of
$h^2(\rho)$ around $\rho=\rho_0$ one finds that  the condition of
stability for  a circular orbit is   
\begin{equation}
h h_{,\rho}  >0,
\end{equation}
or, in other words,
\begin{equation}
h_{,\rho}^2  >0.
\end{equation}

\begin{figure*}
$$
\begin{array}{cc}
v_c^2    &  h^2  \\
\epsfig{width=2.0in,file=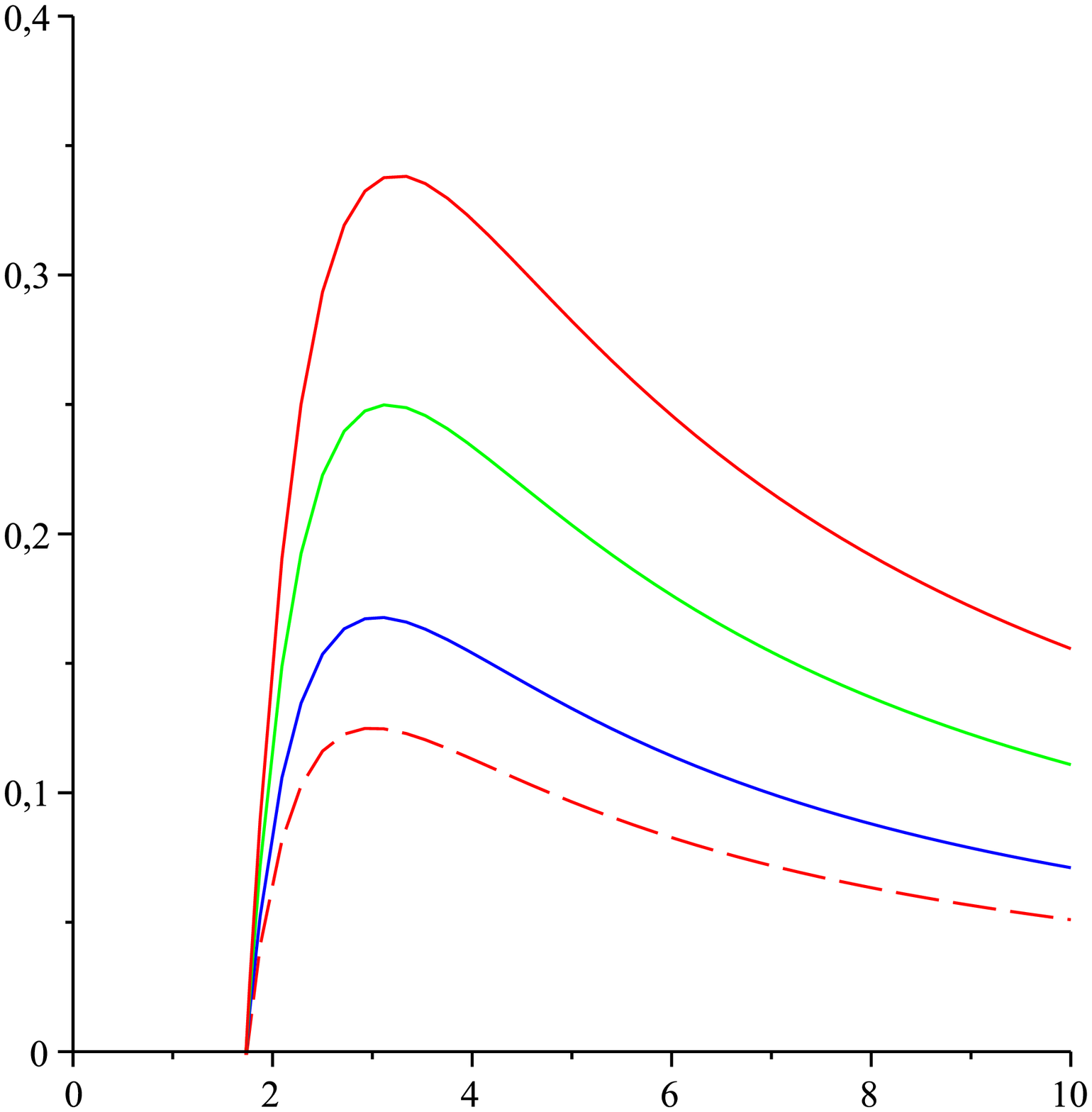} &
\epsfig{width=2.0in,file=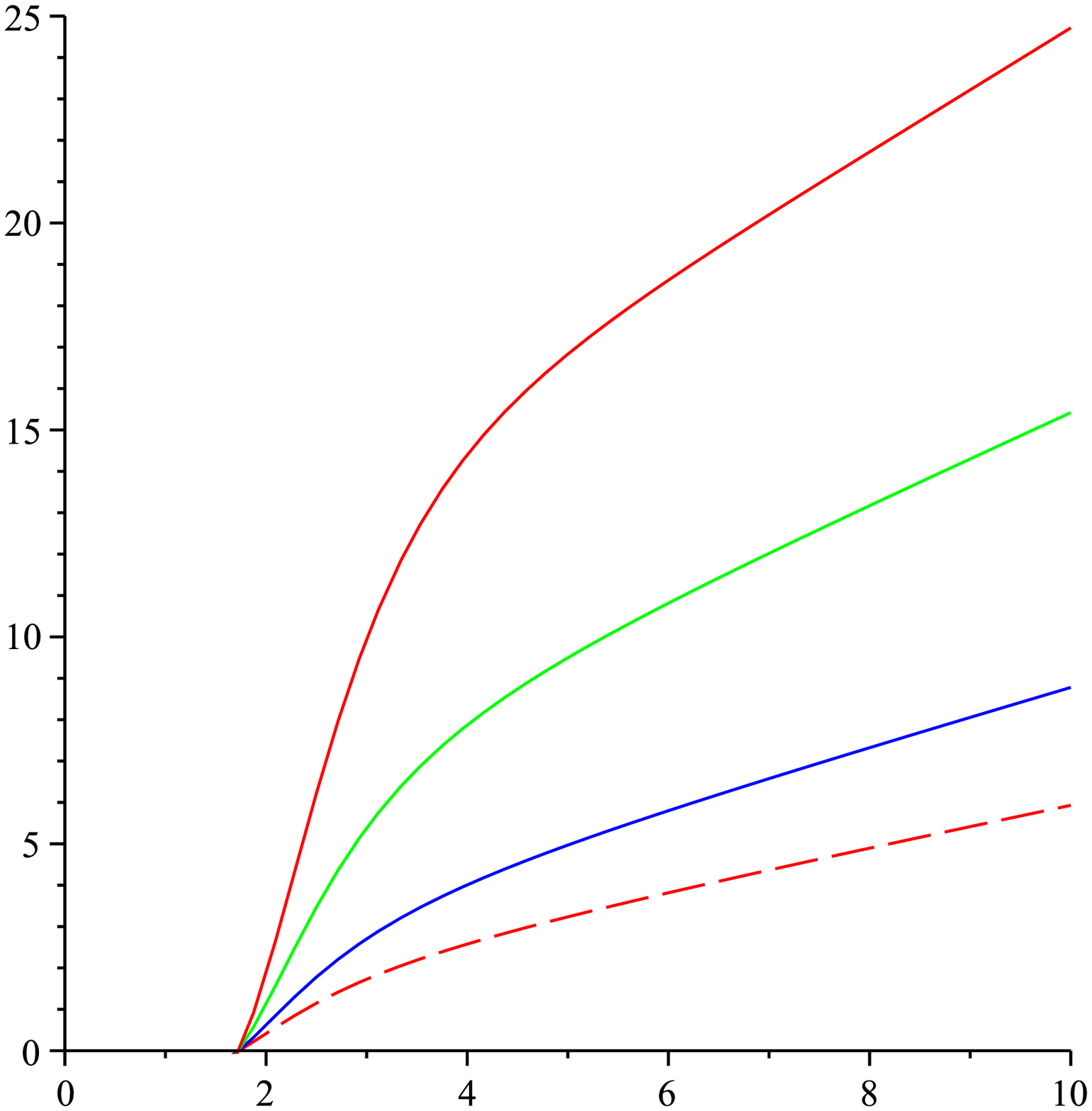}  \\
 \rho &  \rho   \\
(a)     &   (b) \\
 &  \\
v_c^2    &  h^2  \\
\epsfig{width=2.0in,file=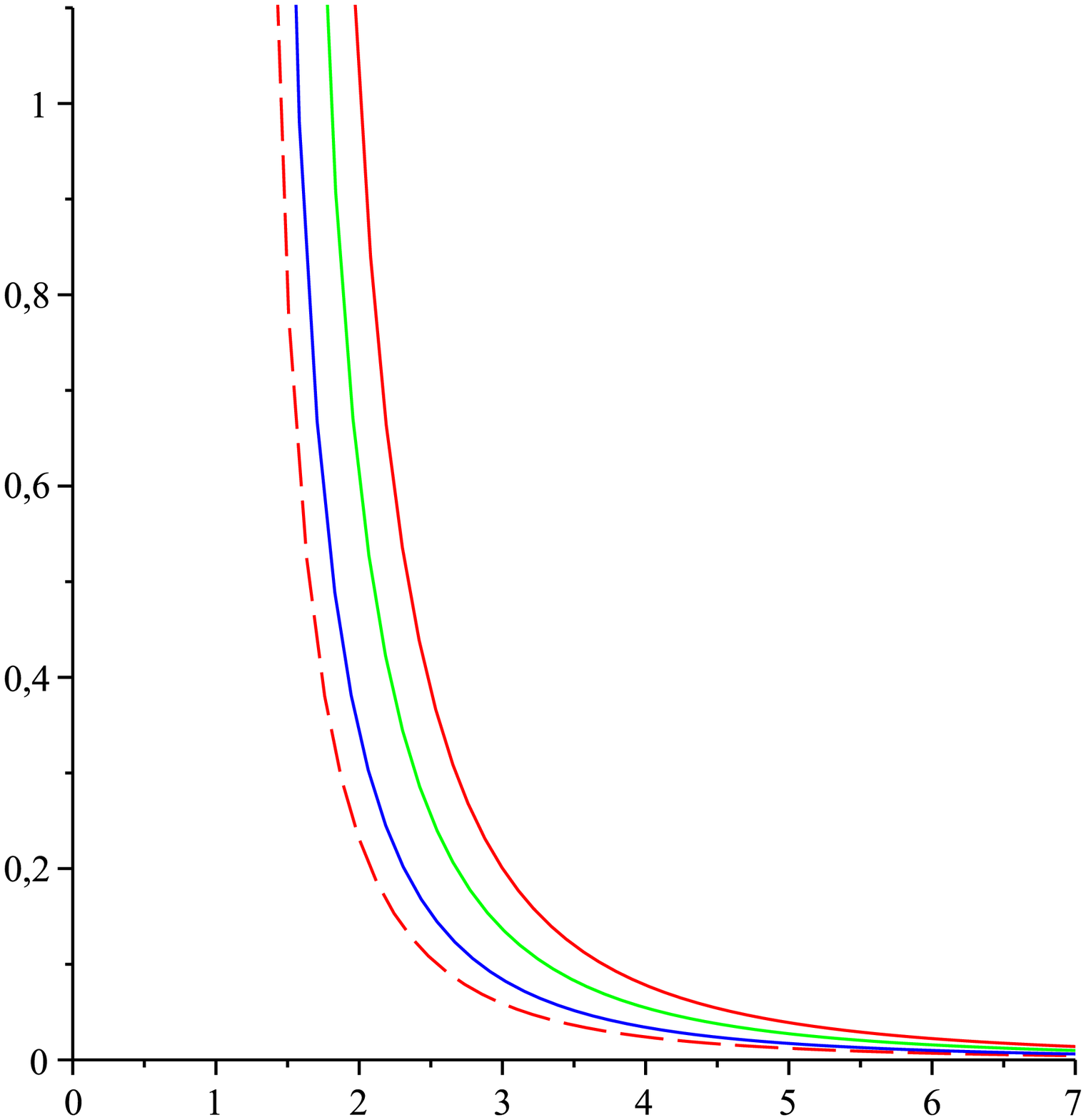} & 
\epsfig{width=2.0in,file=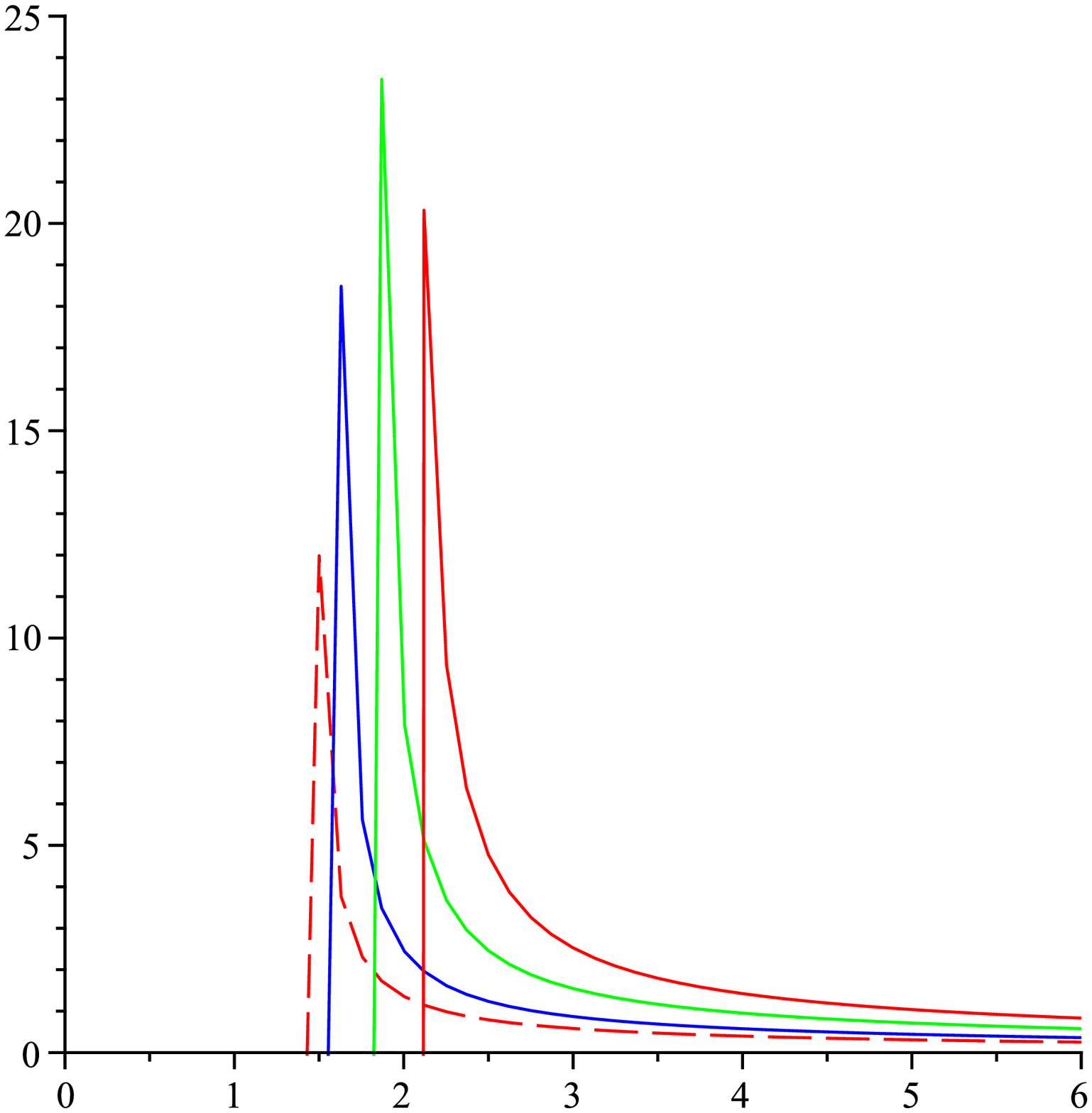} \\
\rho &  \rho    \\
(c)     &   (d) 
 \end{array}
$$	
\caption{For magnetized weyl fields
in spherical  coordinates and second order we plot, as  functions of $\rho$,    the circular velocity  $v_c^2$ and  $h^2$ for test particles
with  $c_0=0.5$ and  $c_2 = 1 $ (figures $(a)$ y $(b)$ ), and 
$c_0=0$ and  $c_2 =-1 $ (figures $(c)$ y $(d)$ ), for  values of magnetic field parameter  $b= 0$ (dashed curves), $1$, $2$, $3$  (top curves).}
\label{fig:esfericas}
\end{figure*}

In order to study the behavior of these physical quantities for the
previous solutions again we perform a graphical analysis of them for  the  zeroth and  second terms of the series in the three families of solutions considered. In all cases we take $\tilde e =1$.  We first analyze the magnetized
Weyl solutions in spherical coordinates. 
In this case  $A_{,\rho}=0$, so that the velocity is given by expression (\ref{eq:v2neutras}).
In Figure \ref{fig:esfericas} we show  
the circular velocity curves  $v_c^2$ and  $h^2$ for test particles
with    $c_0=0.5$ and  $c_2 = 1 $ (figures \ref{fig:esfericas}$(a)$ and  \ref{fig:esfericas}$(b)$), and   $c_0=0$ and  $c_2 =-0.5 $

(figures \ref{fig:esfericas}$(c)$ and  \ref{fig:esfericas}$(d)$), for  values of magnetic field parameter  $b= 0$ (dashed curves), 
$1$, $2$, $3$ (top curves), as  functions of $\rho$.
In first case  we see that  the tangential velocity  increase initially from certain $\rho=\rho_0$, reaches  a maximum and then falls to zero, and is  
always  a quantity less than the velocity of light.  We also observer  that inclusion of magnetic field make  these orbits less relativistic. Meantime  we see that  the  quantity  $h^2$ always is a increasing   function of $\rho$  that corresponds to stable circular orbits.  
In the second case we find that $v^2_c$ increases from zero at infinity to 1 (the velocity of light) at circular photon orbit $\rho=\rho_{ph}$, and   that $h^2$ always is a decreasing monotonous function of $\rho$ which means that this orbits are instable against
radial perturbation.

\begin{figure*}
$$
\begin{array}{cc}
v_c^2    &  h^2  \\
\epsfig{width=2.0in,file=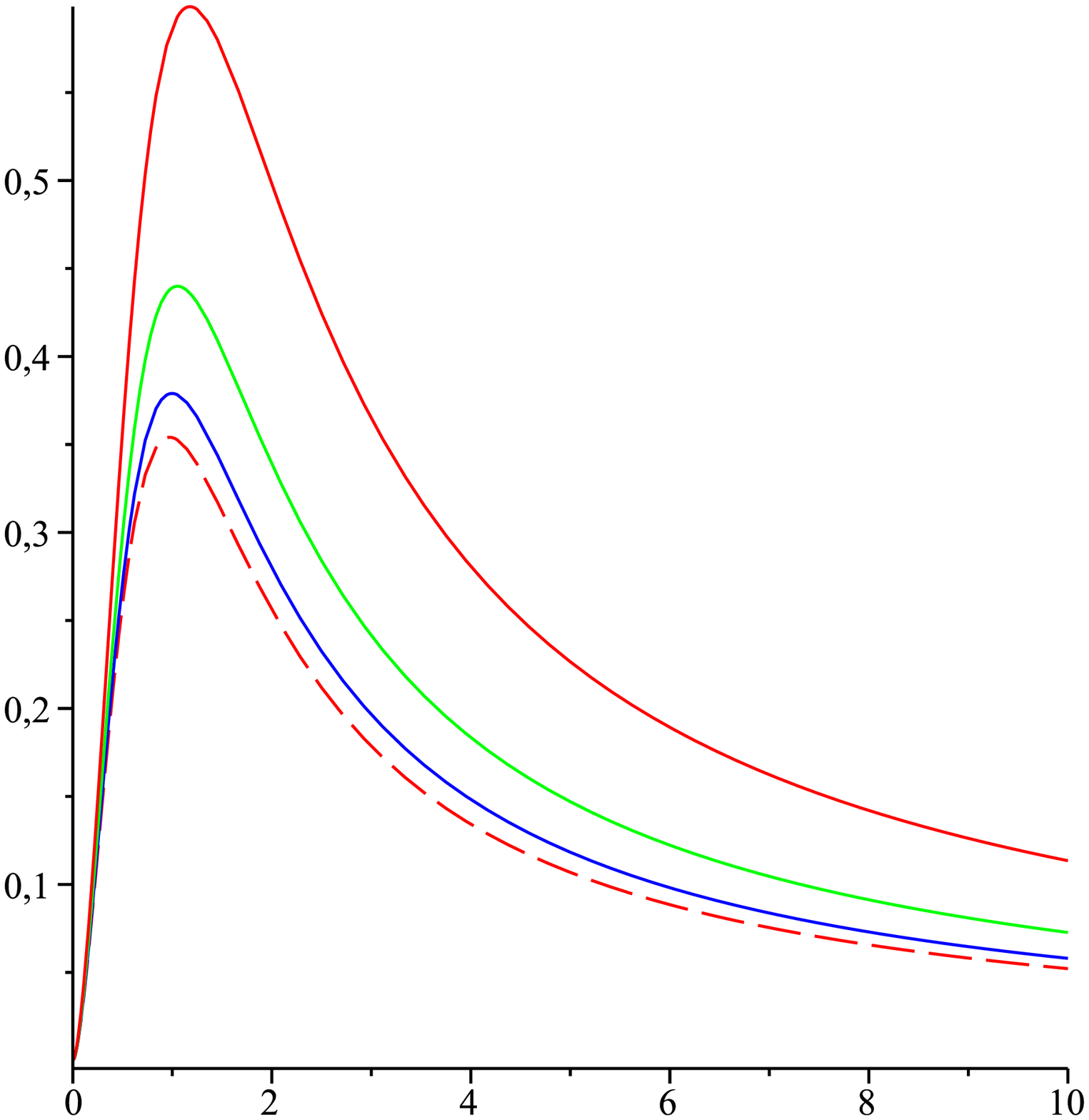} &
\epsfig{width=2.0in,file=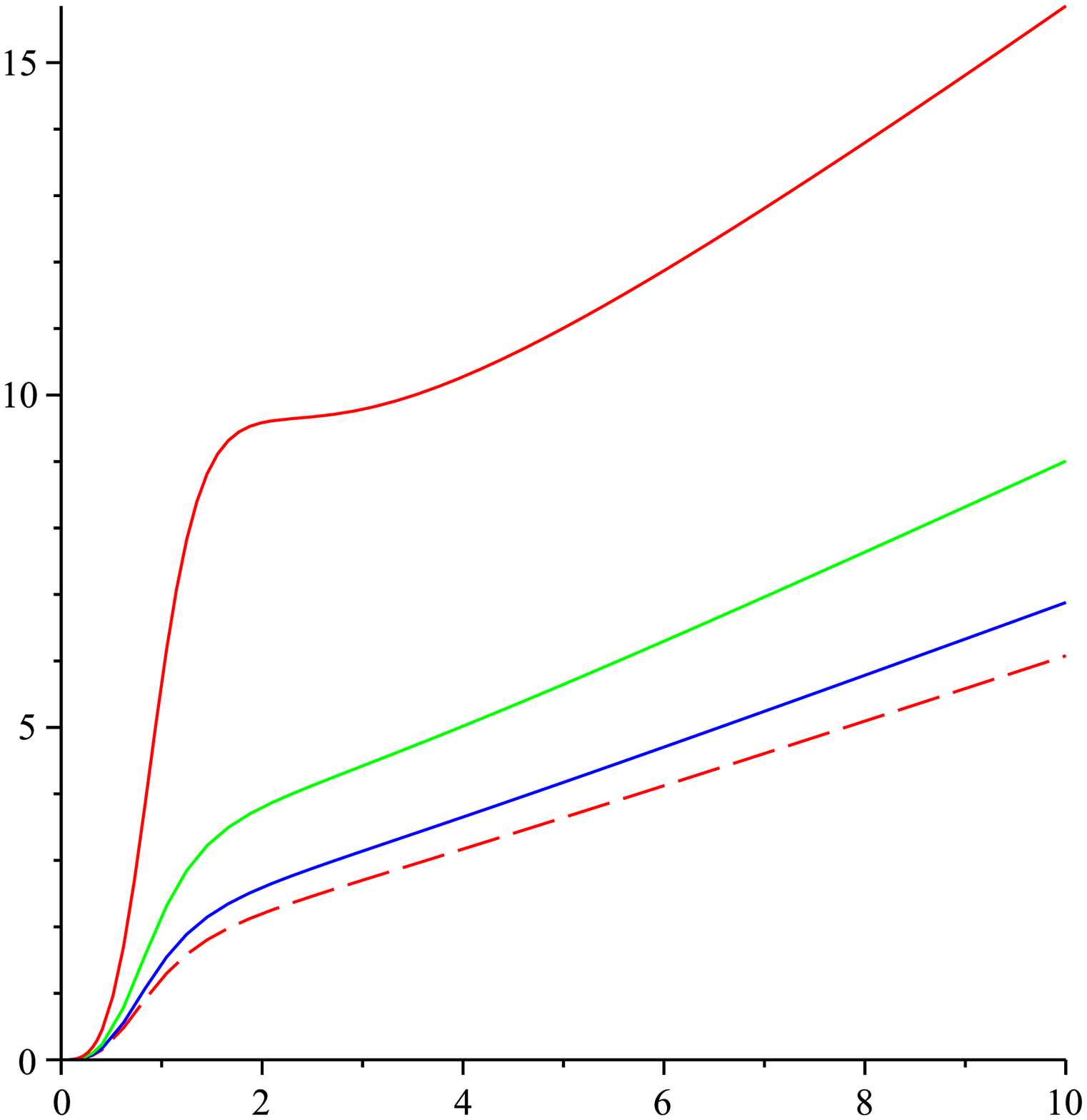}  \\
 \rho &  \rho    \\
(a)     &   (b) \\
 &  \\
v_c^2    &  h^2  \\
\epsfig{width=2.0in,file=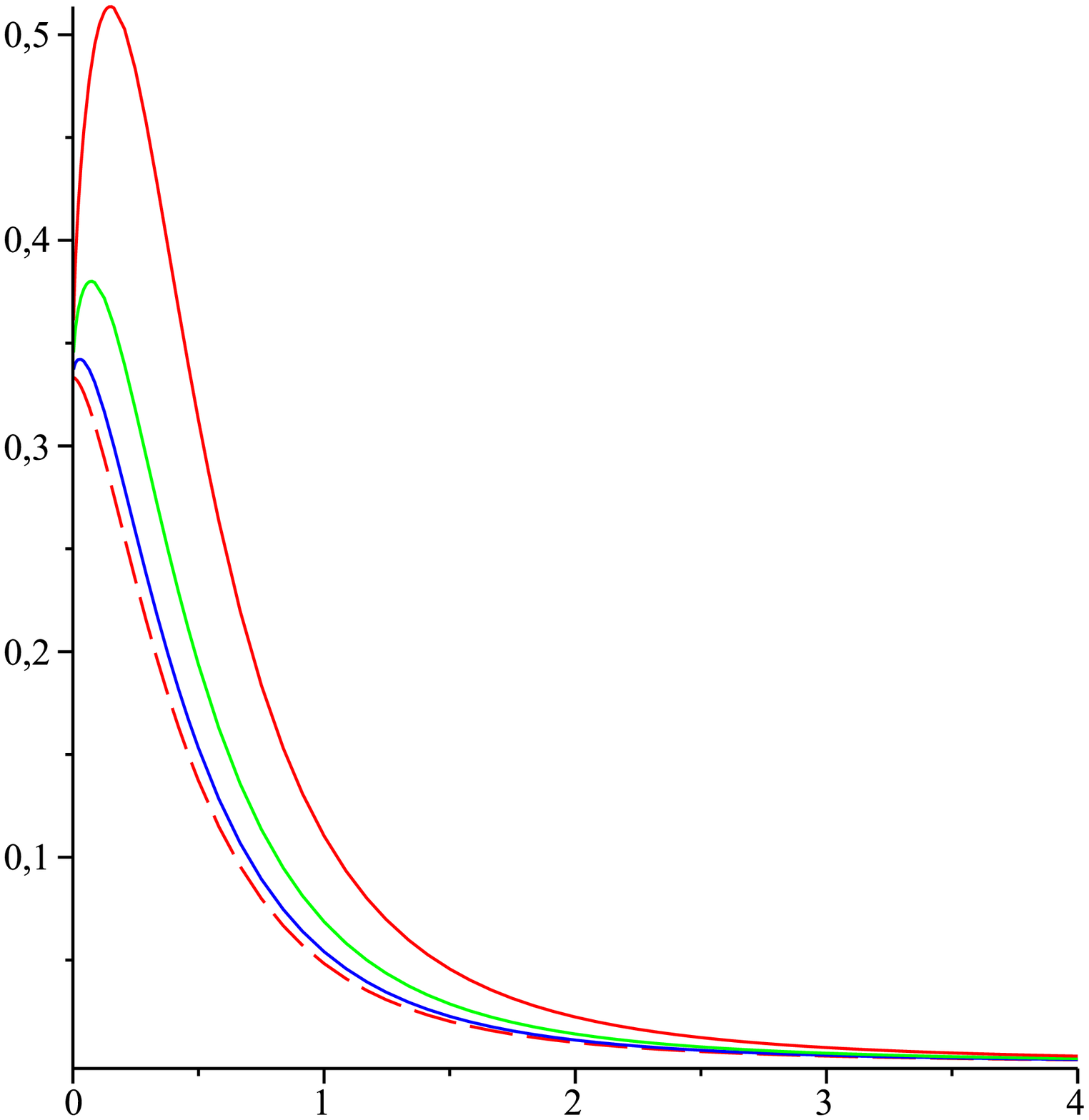} & 
\epsfig{width=2.0in,file=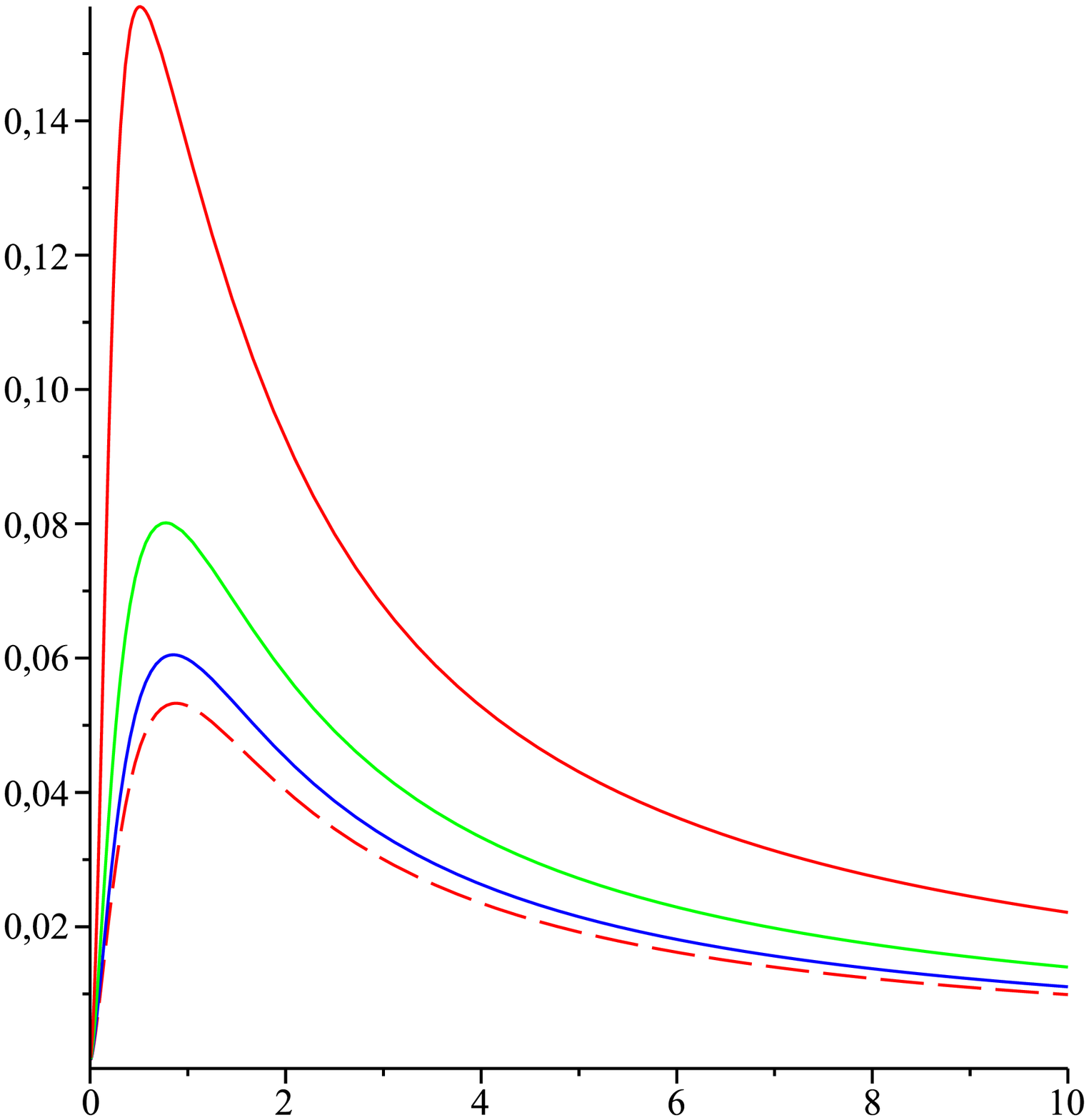} \\
\rho &  \rho    \\
(c)     &   (d) 
 \end{array}
$$	
\caption{For magnetized Erez-Rosen fields
we plot, as  functions of $\rho$,    the circular velocity  $v_c^2$ and  $h^2$ for test particles
with  $c_0=0.5$ and  $c_2 = 1 $ (figures $(a)$ y $(b)$ ), and 
$c_0=0$ and  $c_2 =-0.5 $ (figures $(c)$ y $(d)$ ), for  values of magnetic field parameter  $b= 0$ (dashed curves), $0.5$, $1$, $2$ (top curves).}
\label{fig:erez-rosen}
\end{figure*}

We now analyze the magnetized Erez-Rosen  solutions.  In this case 
also  $A_{,\rho}=0$. So in figure \ref{fig:erez-rosen} we show  
the circular velocity curves  $v_c^2$ and  $h^2$ for test particles
with    $c_0=0.5$ and  $c_2 = 1 $ (figures \ref{fig:erez-rosen}$(a)$ and  \ref{fig:erez-rosen}$(b)$), and   $c_0=0$ and  $c_2 =-0.5 $
(figures \ref{fig:erez-rosen}$(c)$ and  \ref{fig:erez-rosen}$(d)$), for  values of magnetic field parameter  $b= 0$ (dashed curves), $0.5$, $1$, $2$ (top curves), as  functions of $\rho$.
In first case  we see that  the tangential velocity  increase initially, reaches  a maximum and then falls to zero, being 
always  a quantity less than the velocity of light.  We also observer  that inclusion of magnetic field make  these orbits less relativistic.  The quantity  $h^2$ always is a increasing  function of $\rho$  that corresponds to stable circular  orbits. In the second case we find that $v^2_c$ also increase initially  but not from zero and  that   only the central regions are stable.    

\begin{figure*}
$$
\begin{array}{cc}
v_{c+}^2    &  v_{c}^2  \\
\epsfig{width=2.0in,file=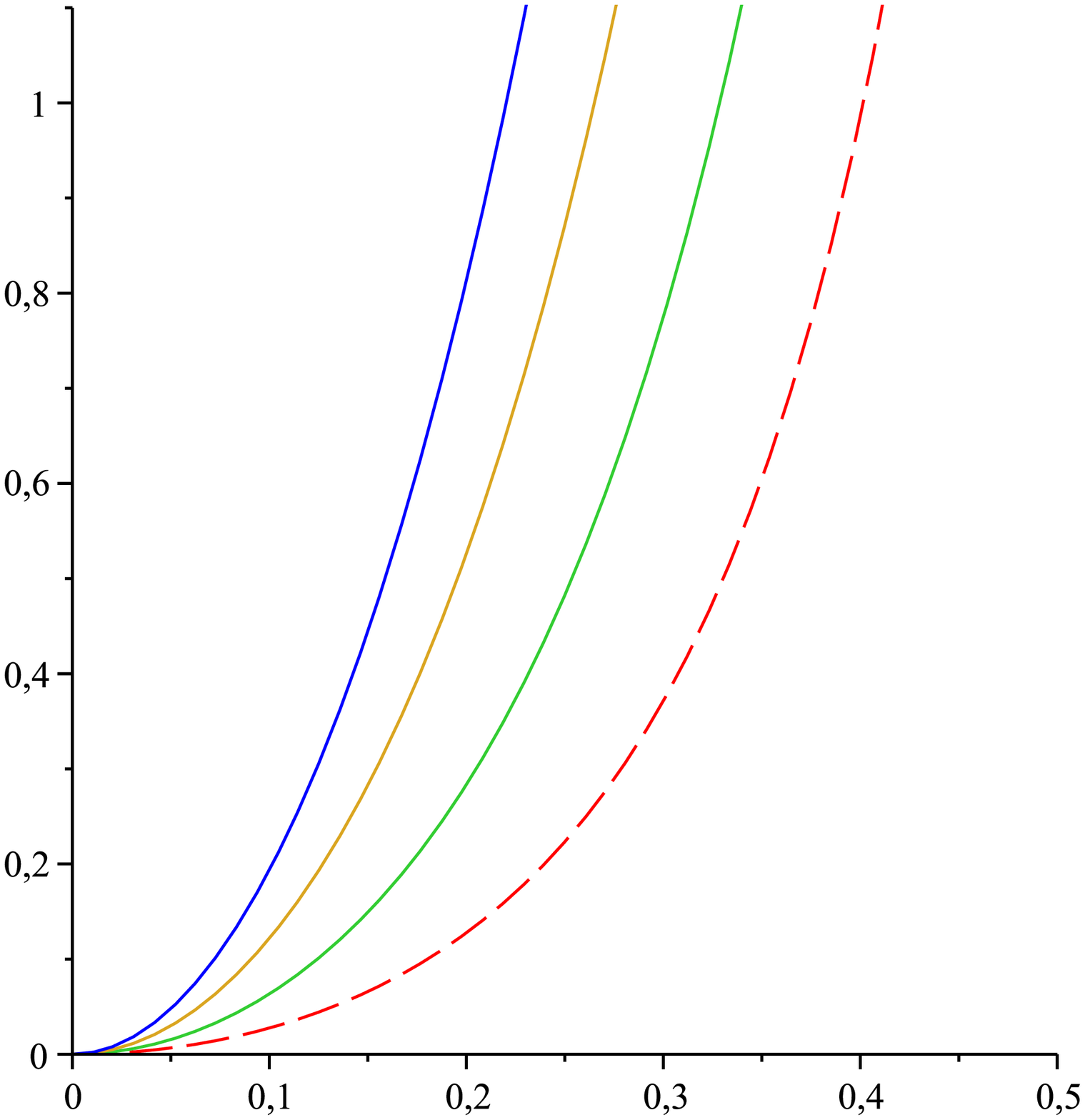} &
\epsfig{width=2.0in,file=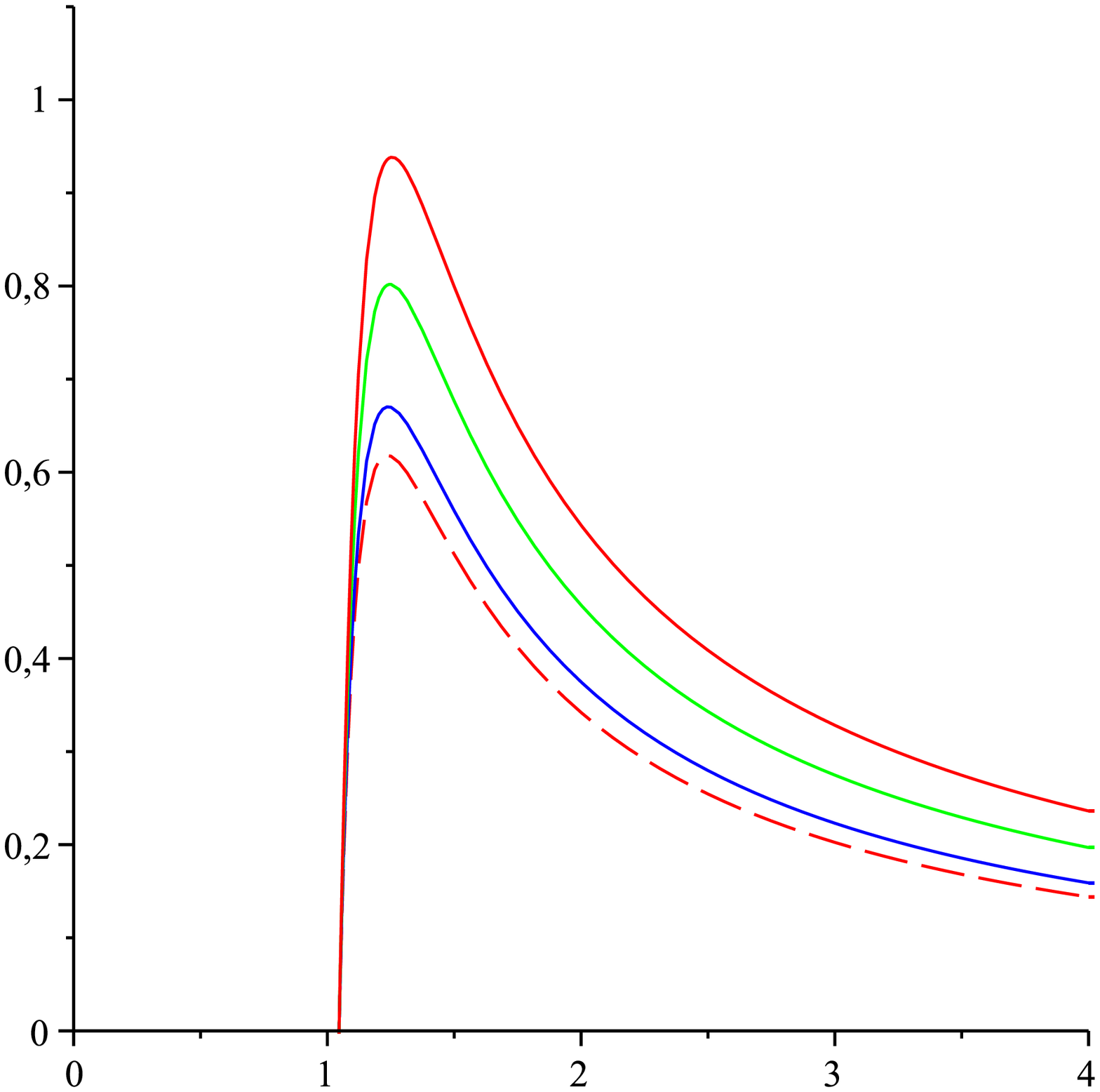}  \\
 \rho &  \rho    \\
(a)     &   (b) \\
 &  \\
h_+^2    &  h^2  \\
\epsfig{width=2.0in,file=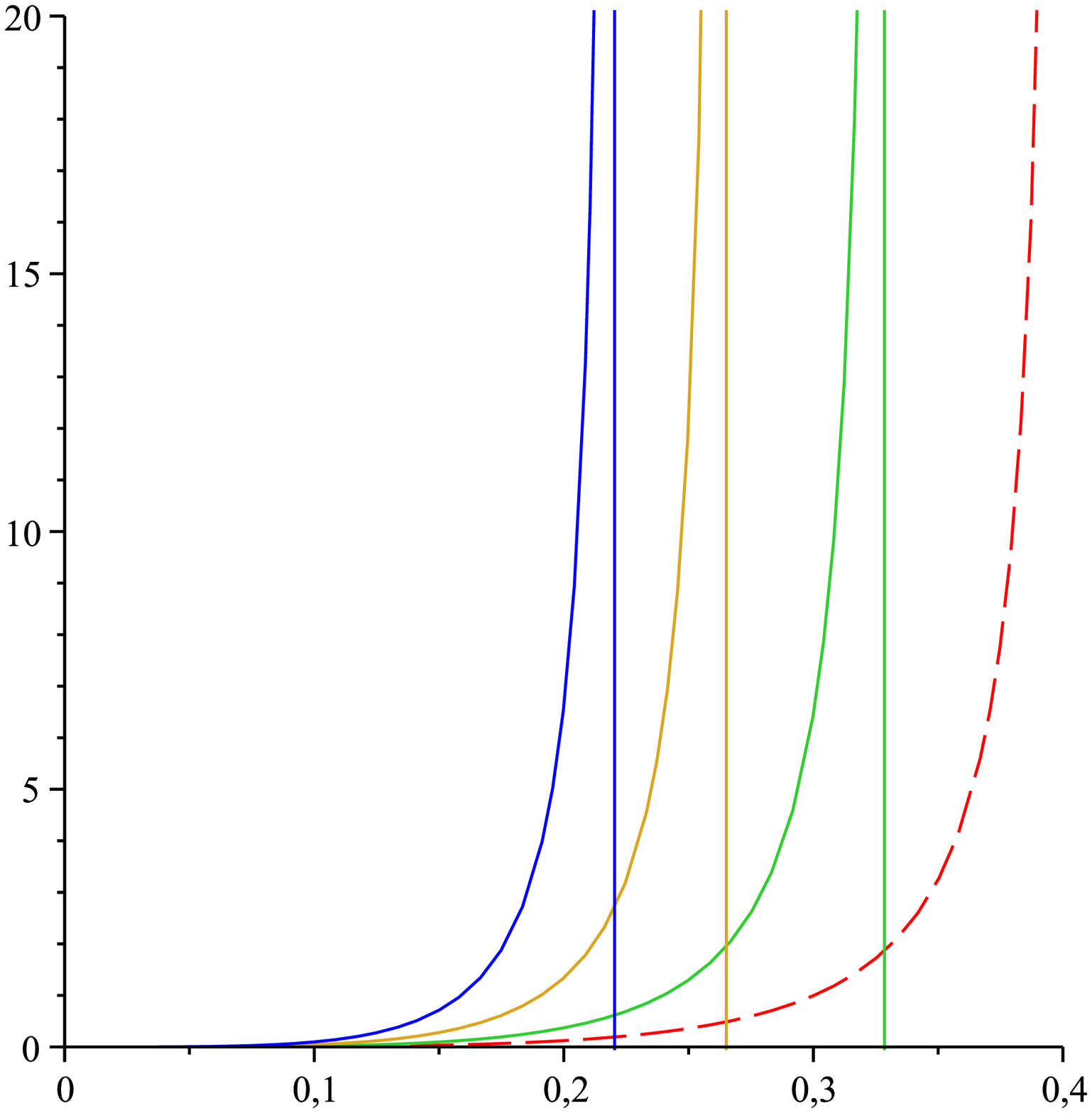} & 
\epsfig{width=2.0in,file=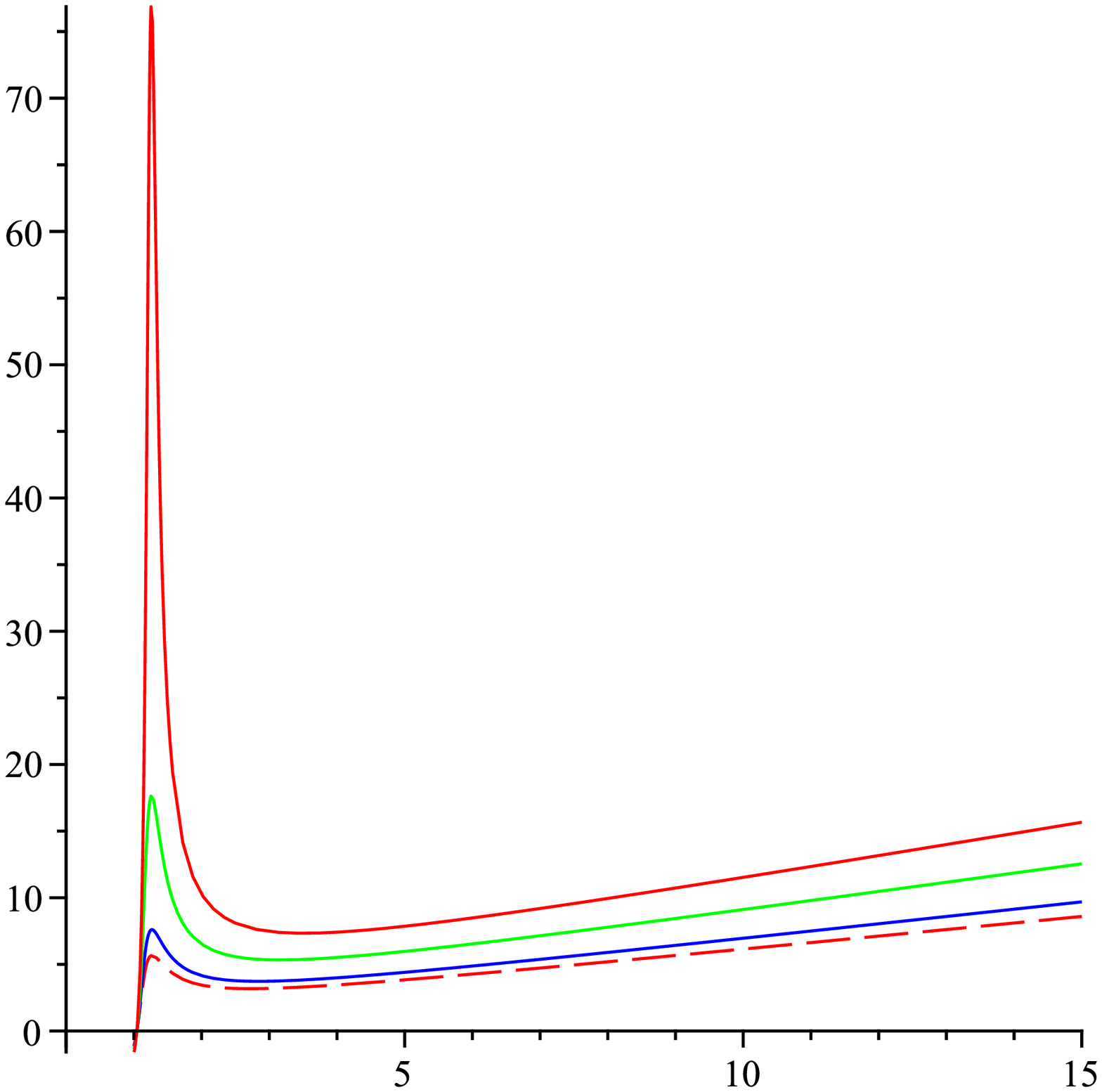} \\
\rho &  \rho    \\
(c)     &   (d) 
 \end{array}
$$	
\caption{For magnetized first order Morgan-Morgan   fields
 we plot,  as  functions of $\rho$,    the circular velocity  $v_{c}^2$ and  $h^2$ for test particles
with  $c_0=0.5$ and  $c_2 = 1 $,  for  values of magnetic field parameter  $b= 0$ (dashed curves), $0.5$, $1$, and $1.4$ (top curves).
Figures on the left side correspond to the inside  of the disk
and to the motion direct of particles.}

\label{fig:morgan1}
\end{figure*}

\begin{figure*}
$$
\begin{array}{cc}
v_{c+}^2    &  v_c^2  \\
\epsfig{width=2.0in,file=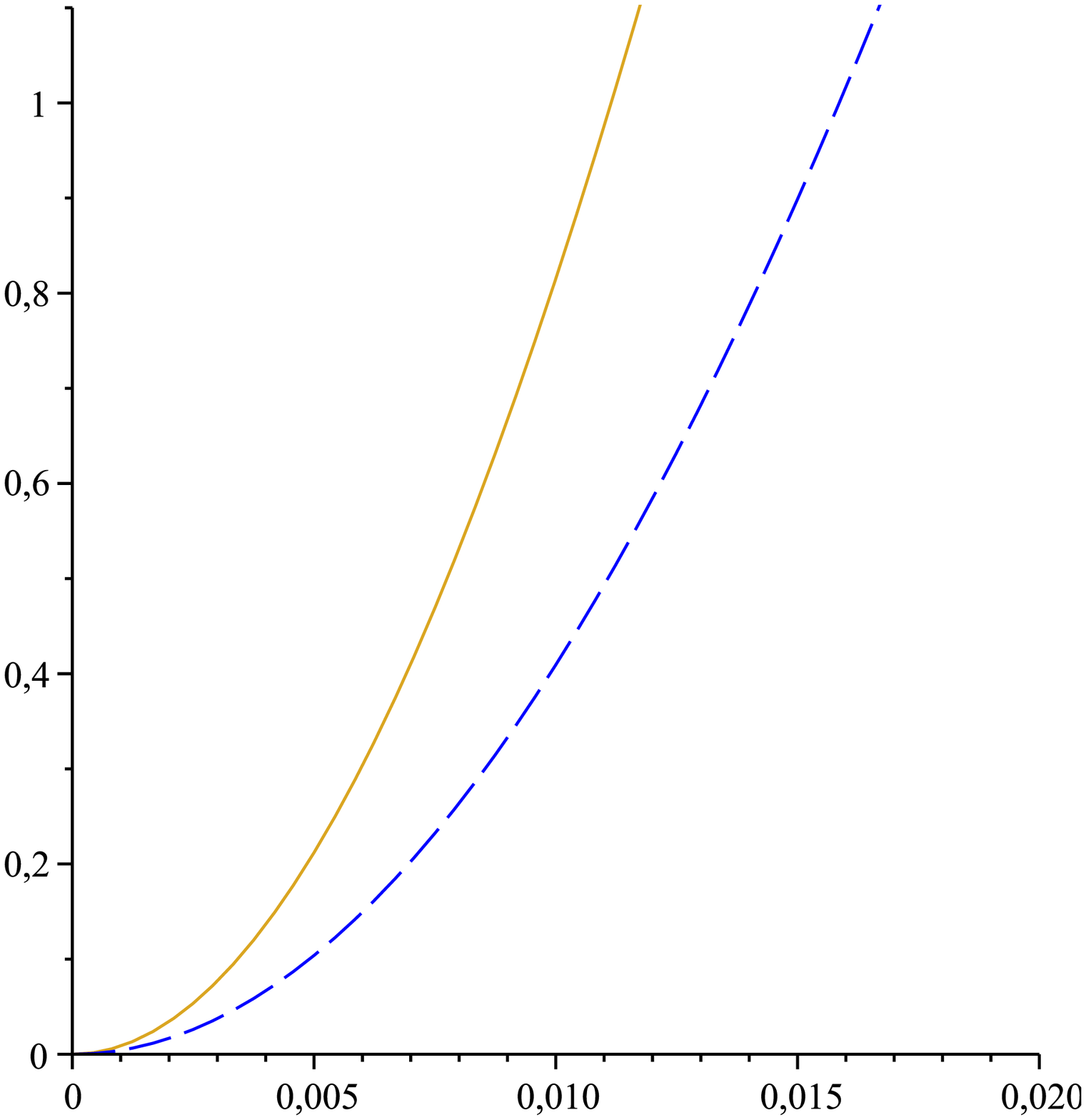} &

\epsfig{width=2.0in,file=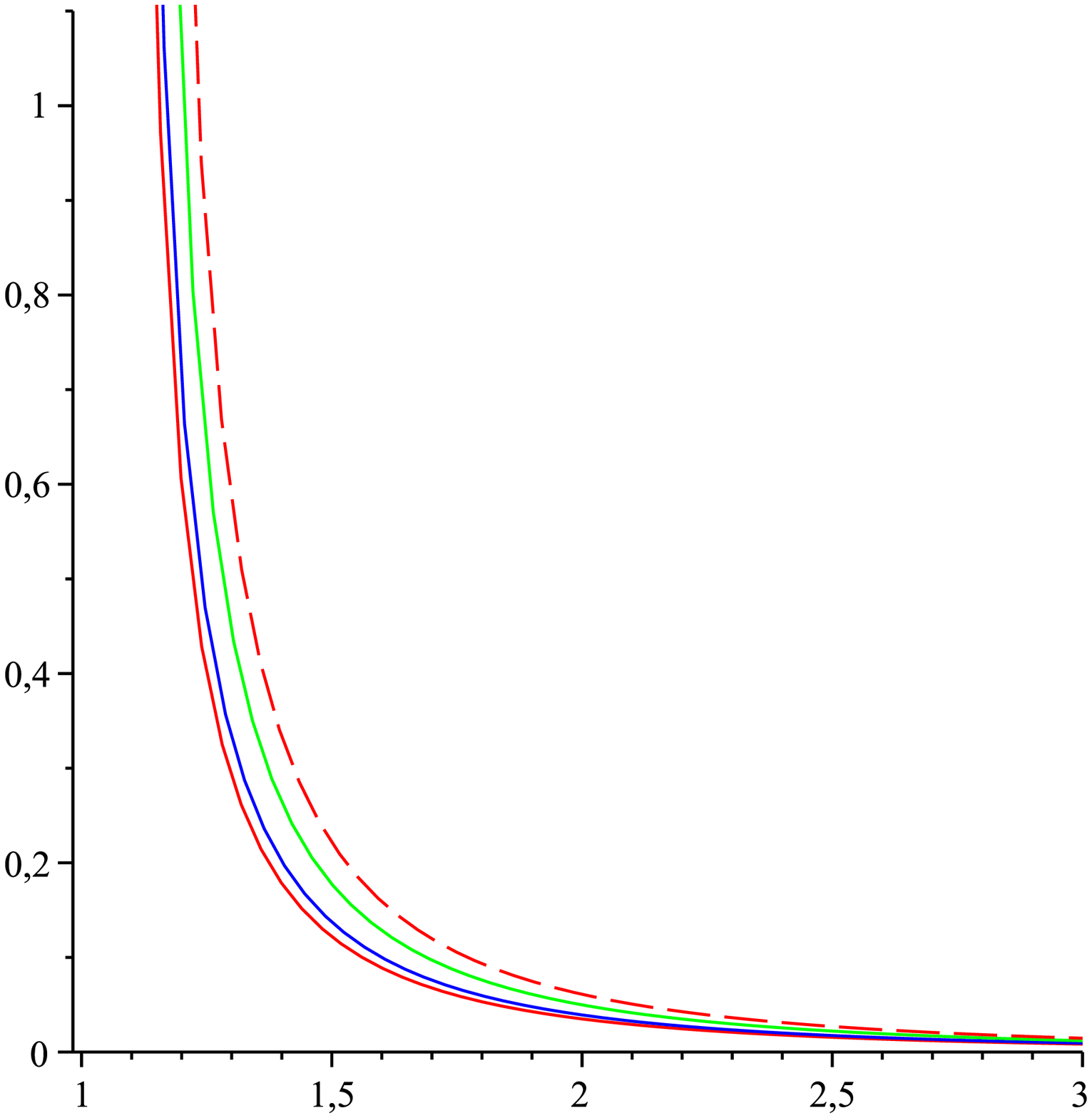}  \\
 \rho &  \rho    \\
(a)     &   (b) \\

 &  \\
h_{+}^2    &  h_+^2  \\

\epsfig{width=2.0in,file=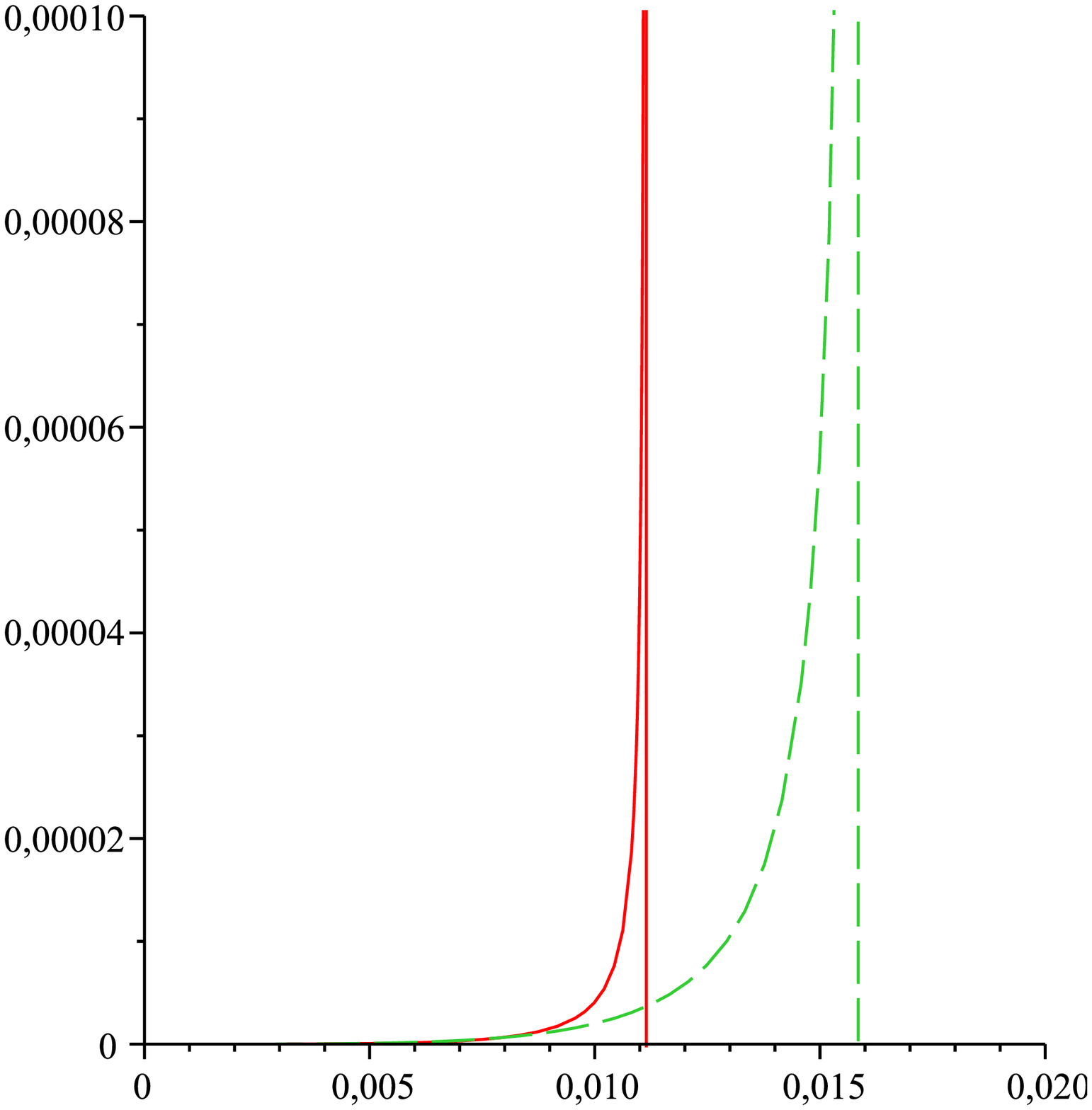} & 
\epsfig{width=2.0in,file=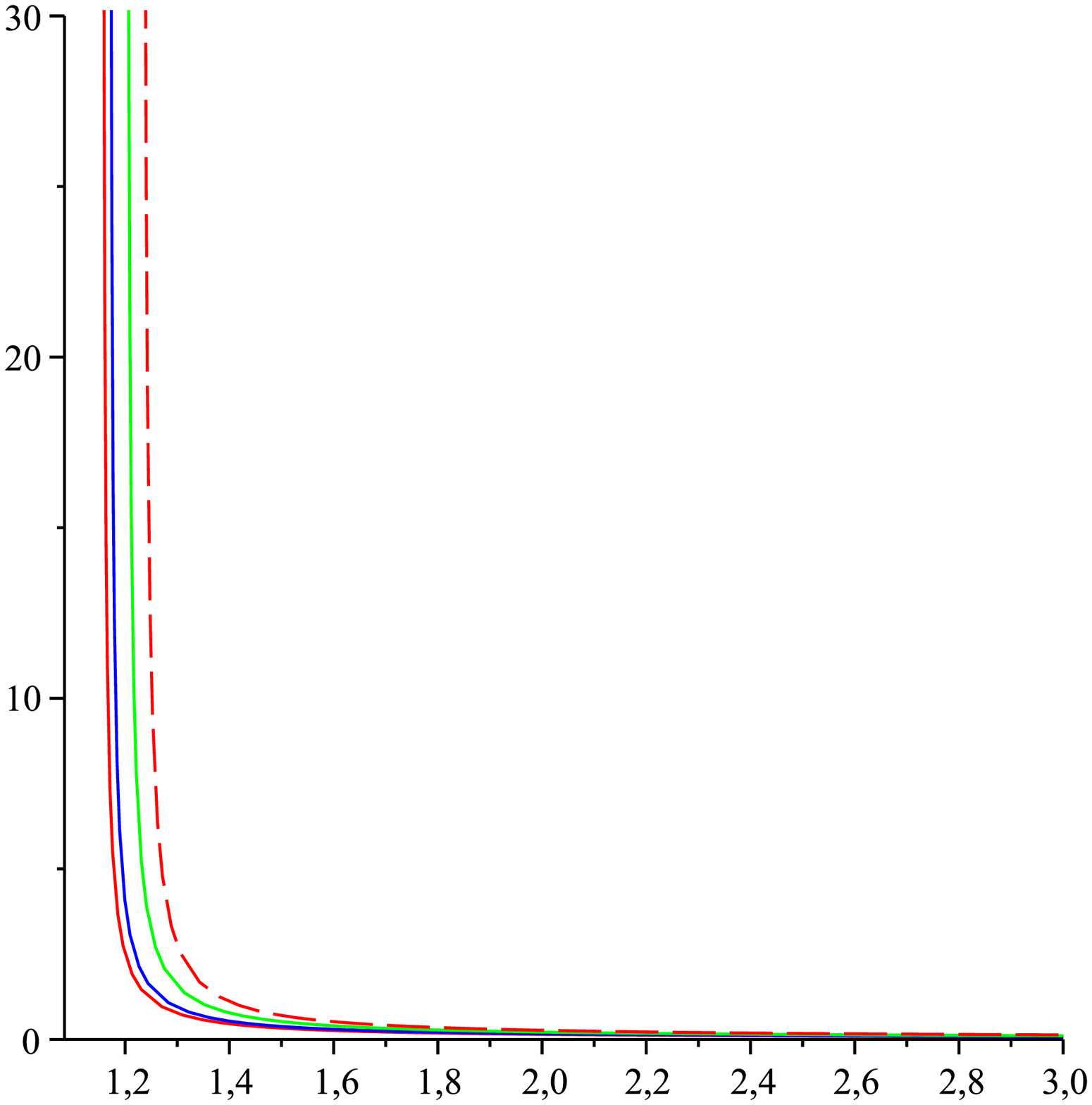} \\
\rho &  \rho   \\

(c)     &   (d) 

 \end{array}
$$	
\caption{Also for magnetized  first order Morgan-Morgan   fields
we plot, as  functions of $\rho$,    the circular velocity  $v_c^2$ and  $h^2$ for test particles
with   $c_0=0$ and  $c_2 =-1$,  for  values of magnetic field parameter 
$b= 0$ , $0.5$, $1$, $1.4$ (dashed curves).  Again figures on the left side correspond to the interior region  of the disk and the motion direct of particles.}
\label{fig:morgan2}
\end{figure*}

In Figures \ref{fig:morgan1} and  \ref{fig:morgan2} we show  
for magnetized  first order Morgan-Morgan   fields
 the circular velocity curves  $v_c^2$ and  $h^2$ for test particles
with    $c_0=0.5$, $c_2 = 1 $  (figures \ref{fig:morgan1})  and  with 
$c_0=0$, $c_2 =-1$  (figures \ref{fig:morgan2}) for  values of magnetic field parameter  $b= 0$, $1$, and $1.4$.
We find inside of the disk $A_{,\rho} \neq 0$ so that in this case we have two values for the angular velocity  $\omega _\pm$, which 
corresponds to the  direct and retrograde motions of particles.
Figures on the left side correspond to the inside  of the disk
and to the direct motion  of particles. For the retrograde motion of the particles we obtain a similar behavior. 
For all values of parameters we find regions where circular orbits
are possible, that is, regions  where the velocity of particles  
is always  a quantity less than the velocity of light.
We also observer  that inclusion of magnetic field can makes  these orbits less relativistic.  
We find strong change in the slope of $h^2$ at certain values of $\rho$  which means that there is a strong instability there, also regions with negative slope where the orbits are unstable, but always we find  regions where $h_{+}^2$ is a increasing monotonous function of $\rho$  that corresponds to stable circular orbits.   We also observe  that the increase of magnetic field can make stable these orbits against radial perturbation.

\begin{figure*}
$$
\begin{array}{ccc}
v_{c+}^2   &   h_{+}^2  &   h_{+}^2   \\
\epsfig{width=2.0in,file=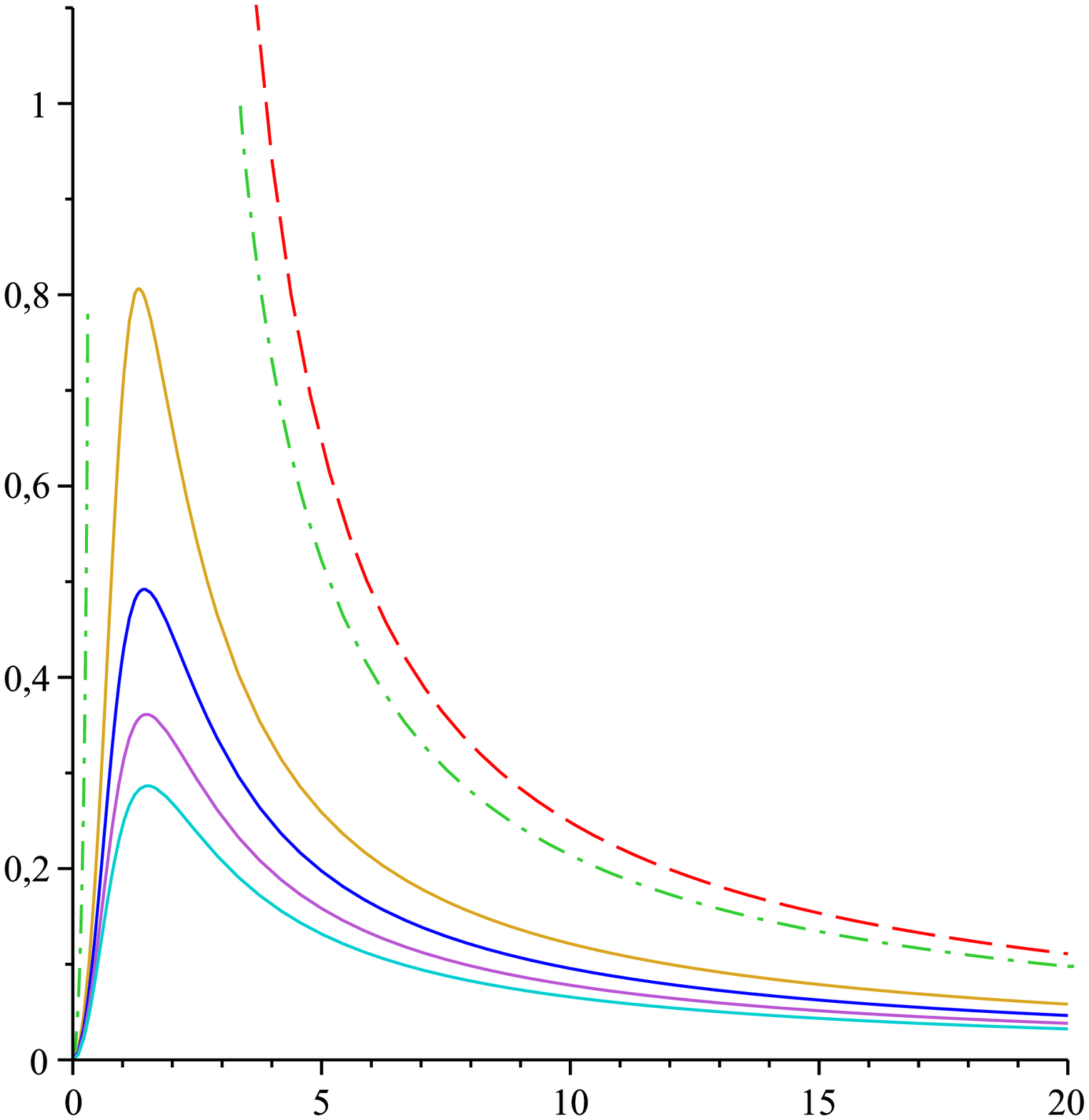} & 
\epsfig{width=2.0in,file=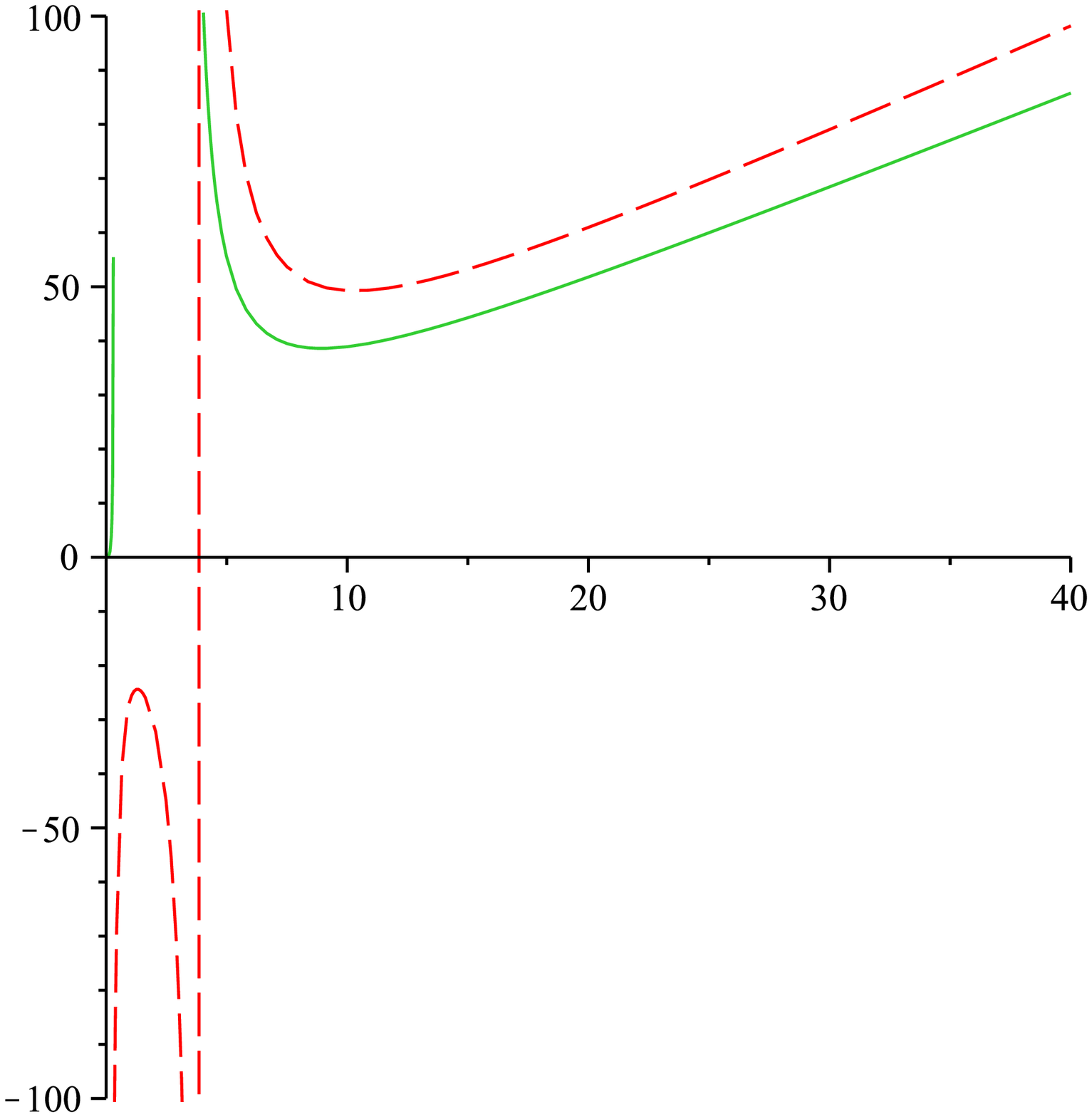}  &
\epsfig{width=2.0in,file=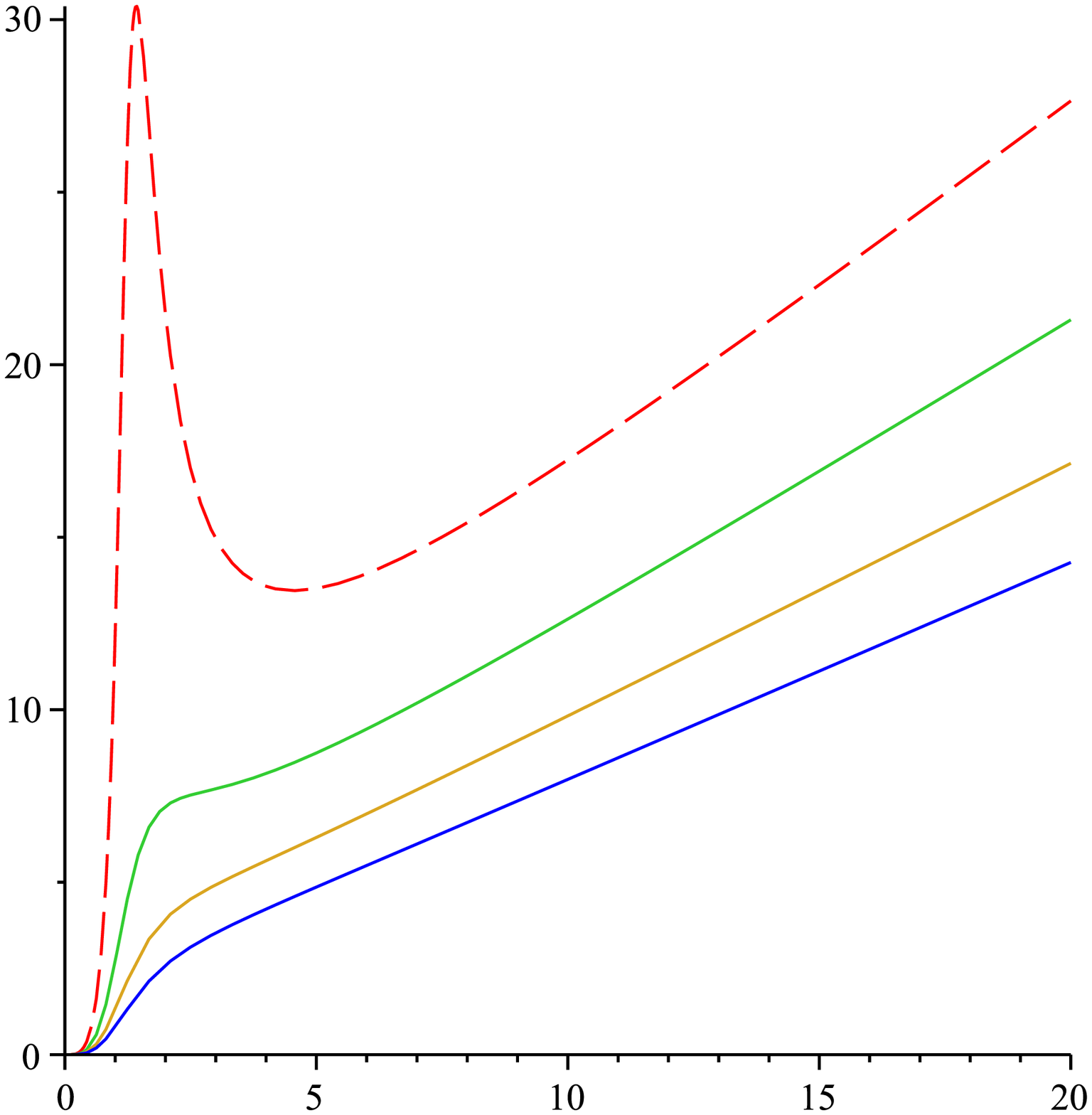} \\
 \rho &  \rho & \rho   \\
(a)     &   (b) &  (c)  \\
\end{array}
$$	
\caption{For  a Kerr-type  solution
(magnetic dipole solution) we plot, as  functions of $\rho$,  $(a)$ the circular velocity  $v_{c+}^2$ for test particles
with  $b= 0$ (dashed curve), $0.5$ (curve with points and lines), $1.5$, $2$,  $2.5$,  and $3$ (bottom curve), and   $h_{+}^2$ for $(b)$ $b=0$  (dashed curve),
$0.5$, $(c)$ $b= 1.5$ (dashed curve), $2$, $2.5$, and $3$ (bottom curve).}
\label{fig:bonnor}
\end{figure*}

In order to compare the behavior of these 
quantities  with other  known  magnetized solutions we now analyze a
Kerr type solution or magnetic dipole solution \cite{BONNOR}.  
In this case also  we have two values for the angular velocity, 
$\omega _\pm$.
For the direct motion of the particles,  in figure \ref{fig:bonnor} we present 
\ref{fig:bonnor}$(a)$ the circular velocity  curves $v_{c+}^2$ 
for $b= 0$ (dashed curve), $0.5$ (curve with points and lines), $1.5$, $2$,  $2.5$,  and $3$ (bottom curve), and   $h_{+}^2$ for \ref{fig:bonnor}$(b)$ $b=0$  (dashed curve),
$0.5$, \ref{fig:bonnor}$(c)$ $b= 1.5$ (dashed curve), $2$, $2.5$, and $3$ (bottom curve), as functions of $\rho$. We see that in absence of magnetic
field $b=0$ the circular velocity  increases from zero at infinity to 1 (the velocity of light) at circular photon orbit $\rho=\rho_{ph}$, but as we increase the magnetic field it reaches  a maximum and then falls to zero, being always less than the velocity of light. We also see that inclusion of magnetic field make  these orbits less relativistic. Regarding  the function $h^2$ we find for   $b=0$ and $b=0.5$ (or weak fields) 
strong change in the slope of $h_{+}^2$ at certain values of $\rho$  which means that there is a strong instability there, also regions with negative slope where the orbits are unstable, but after  certain values of $\rho$ we find that $h_{+}^2$ is a increasing monotonous function of $\rho$  that corresponds to stable circular orbits.   We also observe  that the increase of magnetic field can make stable these orbits against radial perturbation.   For the retrograde motion of the particles we obtain a similar behavior. 


\section{Discussion}

Using the well-known Ernst's method were constructed  three families of  general magnetostatic axisymmetric exact solutions of Einstein-Maxwell equations in spherical coordinates, prolate, oblates, and also in generalized spheroidal coordinates which is a generalization of the previous systems.  Except the zero order, the constructed solutions are asymptotically flat but with  total mass  zero. Hence, all discussed solutions with magnetic field  are either singular, not asymptotically flat or massless.
In the three cases considered we  show   explicitly  the solutions corresponding to  the zeroth and second terms of the series of the solutions,  among them a  magnetized Erez-Rosen metric and
a magnetized  Morgan-Morgan  metric, which was interpreted  as  the exterior gravitational field of  a finite dislike source immersed in a magnetic field, and we analyze the material properties of the disk such as the
surface  energy density and the azimuthal pressure. From them  we 
also study relativistic models of thin disks of infinite extension. We find always values of parameters for which   all the energy conditions are satisfied.  

We also  analyzed  the electrogeodesic  equatorial circular  motion of test particles for the sum of the first and  second terms of series of solutions and we also discuss their stability.
For all values of parameters we find regions where circular orbits
 are possible, that is, regions  where the circular velocity of particles  
is always  a quantity less than the velocity of light.
We also observer  that inclusion of magnetic field can makes  these orbits less relativistic.  
We find strong change in the slope of the function $h^2$ at certain values of the radial coordinate $\rho$  which means that there is a strong instability there, also regions with negative slope where the circular orbits are unstable, but always we find values of parameter  for which $h^2$ is a increasing monotonous function of $\rho$  that corresponds to stable orbits.   We also observe  that the increase of magnetic field can make stable these orbits against radial perturbation.    
For  magnetized  Morgan-Morgan fields was found  that inside of disk  the presence of magnetic field provides the possibility of to find 
relativist charged particles  moving  in both
direct and retrograde direction.

Finally in order to compare the behavior of these 
physical quantities  with other  known  magnetized solutions,  we 
also analyze the electrogeodesic motion of test particles  and  their stability   for a  Kerr-type solution  (magnetic dipole solution). We found for some value of  the parameter a similar behavior to the previous cases.


\newpage





\end{document}